\documentstyle[epsf]{mn}
\begin{document}
\input{epsf}
\title[Axisymmetry in proto-planetary nebulae]
{Axisymmetry in proto-planetary nebulae: using imaging polarimetry to
investigate envelope structure}

\author[T.M. Gledhill et al.]
        {T.M.~Gledhill,$^{1}$\thanks{email: T.Gledhill@star.herts.ac.uk}
         A.~Chrysostomou,$^{1}$
         J.H.~Hough$^{1}$
         and J.A.~Yates$^{1}$ \\
         $^{1}$Department of Physical Sciences, University of
        Hertfordshire, 
        College Lane, Hatfield, \\ Hertfordshire AL10 9AB, England
         }
\maketitle

\begin{abstract}
We use ground-based imaging polarimetry to detect and image the dusty
circumstellar envelopes of a sample of proto-planetary nebulae (PPNe)
at near-infrared wavelengths. This technique allows the scattered
light from the faint envelope to be separated from the glare of the
bright central star and is particularly well suited to this class of
object.  We detect extended (up to $9$ arcsec diameter) circumstellar
envelopes around 15 out of 16 sources with a range of morphologies
including bipolars and shells. The distribution of scattered light in
combination with its polarization (up to 40 per cent) provides
unambiguous evidence for axisymmetry in 14 objects showing this to be
a common trait of PPNe.  We suggest that the range of observed
envelope morphologies results from the development of an axisymmetric
dust distribution during the superwind phase at the end of the AGB. We
identify shells seen in polarized light with scattering from these
superwind dust distributions which allows us to provide constraints on
the duration of the superwind phase. In one object (IRAS 19475+3119)
the circumstellar envelope has a two-armed spiral structure which we 
suggest results from the interaction of the mass losing star with a binary 
companion.
\end{abstract}


\section{Introduction}

Proto-planetary nebulae (PPNe) represent a transition phase in the
evolution of low and intermediate mass stars between the Asymptotic
Giant Branch (AGB) and the Planetary Nebula (PN) phases (Kwok 1993).
During its AGB lifetime, a star undergoes periods of intense mass
loss, with mass loss rates of up to $10^{-5}M_{\sun}$yr$^{-1}$ (Loup et
al. 1993), resulting in the formation of a circumstellar envelope
(CSE) of dust and gas. After mass loss ceases at the end of the AGB,
the envelope continues to expand and to disperse during the next
$\sim10^{3}$yr of evolution as a PPN, and subsequently as a PN.

By studying the CSEs of PPNe much can be learned about the mass loss
process during the AGB phase, since the envelope structure is a
cumulative record of the mass loss history. One particularly striking
development is the emergence of axisymmetry in the CSE, as observed,
for example, in bipolar PPNe such as AFGL 2668 (Sahai et al. 1998a,b),
IRAS 17150-3224 (Kwok, Su \& Hrivnak 1998) and IRAS 17441-2445 (Su et
al. 1998).  It is currently thought that a dramatic increase in mass
loss occurs at the end of the AGB (the `superwind' phase) and that this 
may be responsible for initiating the shift from
spherical to axisymmetric structure in the CSE, if the wind is
intrinsically axisymmetric (e.g. Meixner et al. 1997). The result is
the ejection of a superwind shell of gas and dust which has a
flattened density distribution, being more equatorially concentrated,
which can then collimate any further outflow along the polar axis.

Until recently, however, resolved images of CSEs existed for only a
handful of large, bipolar PPNe, compared with a list of PPNe
candidates numbering over 100 (Volk \& Kwok 1989), and so this model
has not been well tested. In many PPNe, the dust shell has thinned to
the extent that the central star is visible at optical and
near-infrared (NIR) wavelengths which can make them very difficult to
resolve from the ground. If the nebulosity is less than $\sim5$ arcsec
in extent then, with a 4 m class telescope, the CSE is likely to be
hidden by the extended wings of the point spread function (PSF) of the
bright central star, so that the PPN appears star-like from the
ground. To glean any information on the CSE structure requires
deconvolution or PSF subtraction which can be difficult given the huge
difference in the brightness of the central star and the faint CSE
(e.g. Hrivnak et al. 1999a). By moving to space-based observations,
the problems caused by the scattering of light from bright central
stars in the Earth's atmosphere are overcome, allowing the possibiliy
of resolving the CSEs of these PPNe in more detail. Ueta, Meixner \&
Bobrowsky (2000) have undertaken an optical imaging survey of PPNe
using the Wide Field and Planetary Camera 2 (WFPC2) on HST and detect
extended nebulosity in 21 of their 27 targets.

Ground-based imaging polarimetry provides an alternative technique,
which is particularly well suited to resolving structure in the CSEs
of PPNe when they are associated with bright central stars.  Since
PPNe are seen by scattered light in the optical and NIR, they are
expected to be linearly polarized at these wavelengths. On the other
hand, the stellar PSF, which is formed by scattering through tiny
angles in the atmosphere, is effectively unpolarized. Imaging
polarimetry provides the means to separate one from the other so that
in polarized light the central star effectively disappears revealing
the faint CSE. In this paper we use NIR imaging polarimetry to detect
CSEs in a sample of 16 PPNe candidates and to image their structure. 

\section{Experimental Details}

\subsection{Target List}

We have selected targets from the literature which have properties
typical of PPNe and which appear in previously published lists of PPNe
candidates (e.g. Kwok 1993, Meixner et al. 1999). Most have a
mid-infrared (MIR) excess typical of emission from a cool dusty CSE
and a double-peaked SED indicating that the envelope is detatched from
the star. We have not selected the targets on the grounds of
morphology and indeed all appear point source-like in previously
published ground-based imaging. Despite their common properties it is
likely that at least one of our targets, IRAS 18184-1623, is not a PPN
and there is debate over the status of another (IRAS
19114+0002). Nevertheless they are retained in the sample. An
effective cut off of K$>4.2$ was imposed by the NIR detector, below
which targets were too bright to observe without saturation. The target
list is shown in Table 1. The coordinates given are those found in the
{\small SIMBAD} database.

\subsection{Observations}

Observations were made at the 3.8m United Kingdom Infrared Telescope
(UKIRT) on Mauna Kea, Hawaii, using the facility infrared camera
IRCAM3. The 256x256 element InSb detector is sensitive between 1 and 5
microns and was used in conjunction with J, H and K broadband
filters. A warm 2 times magnifying lens was inserted in front of the
cryostat window to provide a scale of $0.143$ arcsec per pixel
with a field of view of approximately 40 arcsec.

Dual beam imaging polarimetry was obtained using an upstream rotating
half waveplate assembly (IRPOL2), along with a
lithium niobate Wollaston prism located in a filter wheel in the
detector cryostat.  An aperture mask mounted at a focal plane between
IRPOL2 and IRCAM3 allows two $40\times10$ arcsec sky segments to be imaged
side by side on the detector in orthogonal polarization states with
minimum overlap. This standard configuration for dual-beam polarimetry
is described in more detail by Berry \& Gledhill (1999).

The data presented in this paper were obtained on two separate
observing runs, in 1998 May and in 1999 June. The 1998 run consisted
of 4 consecutive second half shifts on May 5, 6, 7 and 8 and the run
in 1999 consisted of a single night on June 26.  Conditions on
1998 May 5, 6 and 8 were less than ideal, with varying degrees of
cloud cover. The nights of 1998 May 7 and 1999 June 26 were clear and
photometric. Image correction was provided by a tip-tilt secondary
which resulted in image quality consistently better than 1 arcsec and
frequently better than $0.5$ arcsec (FWHM).  Although an identical
instrument configuration and observing method were used on both runs,
an upgrade to the telescope secondary mirror resulted in a
significantly improved PSF in the 1999 June data.

A total of 16 targets were observed and the details are given in Table
1. The duration of exposures was chosen so as to operate the detector
within its linear response range. Where this was not possible (due to
the brightness of the target) a known linearity correction has been
applied. For exposure times of 1 second and greater a non-destructive
readout mode was used (ND+STARE). For shorter exposures a direct
readout mode (STARE) was used. For extremely bright targets (K$<5$)
requiring exposure times of less than 120 ms, a fast readout
mode was also employed. In all cases the total integration time
includes measurements at several positions on the array (jitters) to
minimize the effect of detector defects. Each jitter position was
observed at four orientations of the half waveplate (separated by 
$22.5\degr$) to obtain linear polarimetry and these images are in turn 
made up of co-added exposures to increase signal to noise.

\subsection{Data Reduction}

After dark subtraction and flatfielding, all target data were reduced
in an identical manner using the dual beam imaging polarimetry package
{\small POLPACK} (Berry \& Gledhill 1999). The software aligns
component images to a common coordinate system and, after correction
for instrumental polarization, combines the data to form resultant
$I$, $Q$ and $U$ Stokes images, from which the other polarized
quantities are obtained. Variance estimates from the raw data are
propagated through the calculation to provide errors on the final $I$,
$Q$ and $U$ images and on the derived polarized quantities $P$ (degree
of polarization) and $I_{p}$ (polarized flux). In all cases the linear
polarization shown has been debiased using the variance on the Q and U
Stokes intensities to account for the non-symmetric noise statistics
produced when adding and squaring Q and U. 

\begin{table*}
\begin{minipage}{150mm}
\caption{A summary of the observations indicating the filters used,
the date of the observation and the exposure and total integration times
in seconds. The full width half maxima (FWHM) of the target and a
nearby reference star, where available, are given for those sources that 
appear point source-like in total flux (all except 20028+3910).}
\label{obs-summary}
\begin{tabular}{|l|c|c|c|c|c|c|c|c|}
IRAS ID     & RA (J2000)   & DEC (J2000)  & Filter & Date & Exposure
            & Integration  & FWHM(target)       & FWHM(ref)       \\
17106-3046  & 17 13 51.7   & -30 49 40    & J  &  26/6/99   & 4  & 432  &0.98 & 0.91\\
            &              &              & K  &  26/6/99   & 1  & 432  &0.74 & 0.68\\ 
17245-3951  & 17 28 04.6   & -39 53 43    & J  &  26/6/99   & 10 & 720  &1.16 &0.95\\
17436+5003  & 17 44 54.9   & +50 02 38    & J  &  5/5/98  & 0.25 & 400  & 0.57 & - \\
            &              &              & K  &  5/5/98  & 0.25 & 300  & 0.50 &0.47 \\
18095+2704  & 18 11 30.6   & +27 05 14    & J  &  5/5/98    & 3 & 540  &1.01 &0.80 \\
            &              &              & K  &  7/5/98 & 0.15  & 180 &0.51 &0.32 \\
18184-1623  & 18 21 19.5   & -16 22 26    & J  &  26/6/99 & 0.072 & 518 &0.72 &0.64 \\
19114+0002  & 19 13 58.7   & +00 07 31    & J  &  5/5/98  & 0.12 & 360  &0.78 &0.81\\
            &              &              & H  &  8/5/98  & 0.12 & 240  &0.85 &0.69\\
            &              &              & K  &  6/5/98  & 0.12 & 360  &0.96 &0.91 \\
19454+2920  & 19 47 24.3   & +29 28 12    & J  &  7/5/98  & 10 & 400   &0.64&0.61\\
            &              &              & K  &  8/5/98    & 2  & 400 &0.55&0.57\\
19475+3119  & 19 49 29.4   & +31 27 15    & J  &  7/5/98  & 0.5 & 200  &0.60 &0.56  \\
            &              &              & K  &  8/5/98  & 0.4 & 320  &0.60 &0.57  \\
19477+2401  & 19 49 54.5   & +24 08 51    & J  &  7/5/98  & 20 & 400  &0.50 &0.50 \\
19500-1709  & 19 52 53.5   & -17 01 50    & J  &  7/5/98  & 0.5 & 200  &0.90 &0.90 \\
            &              &              & K  &  6/5/98  & 0.3 & 240  &0.82 &- \\
20000+3239  & 20 01 59.5   & +32 47 32    & J  &  6/5/98  & 1 & 200  &1.00 &0.94 \\
            &              &              & K  &  7/5/98  & 0.12 & 240  &0.49 &0.50 \\
20028+3910  & 20 04 35.0   & +39 18 38    & J  &  7/5/98  & 20 & 800  & & \\
            &              &              & K  &  6/5/98  & 5 & 400  & & \\
20056+1834  & 20 07 54.8   & +18 42 57    & J  &  5/5/98  & 3 & 504  &0.67 &0.61 \\
            &              &              & H  &  8/5/98  & 1 & 400  &0.61&0.68 \\
21027+5309  & 21 04 14.9   & +53 21 03    & J  &  8/5/98  & 3  & 360 &0.73 &0.73 \\
            &              &              & H  &  8/5/98  & 0.5 & 400 &0.68 &0.68 \\
22223+4327  & 22 24 30.7   & +43 43 03    & J  &  26/6/99 & 0.5 & 680 &0.72 &0.72  \\
22272+5435  & 22 29 10.2   & +54 51 04    & J   & 26/6/99 & 0.072 & 461 &0.80 &0.72 \\
\end{tabular}
\end{minipage}
\end{table*}

\section{Results}

%

\subsection{Format of Results}

In this section we present our imaging polarimetry results, which are
displayed in the same format for each target. In each case we show
greyscale images of total flux superimposed with a polarization vector
map. The polarization vectors are oriented parallel to the E vector
with their length proportional to the degree of linear polarization,
as indicated by the scale vector on each diagram. Unless otherwise
indicated, each polarization measurement (vector) corresponds to an
average over a $3\times3$ pixel ($0.43$ arcsec square) bin. In
addition, a 2$\sigma$ cut has been imposed so that only measurements
with an error in the degree of polarization less than one half of the
polarization are plotted; this has the effect of excluding noisy
measurements in low signal-to-noise regions, such as on the sky.

We also show greyscale images of polarized flux, $I_{p}$ for each
object. Contours are superimposed on the greyscale
images where these serve to highlight structure in the polarized
flux. In these cases the contours are spaced logarithmically at
intervals of 1.0 magnitudes, unless otherwise stated. Since conditions
were non-photometric for most of the run, the data have not been flux
calibrated and the contour levels are therefore in arbitrary
units. The lack of photometric calibration or conditions in no way
affects the polarimetric results since dual beam polarimetry
elliminates the effects of varying sky conditions.

We summarize the results for each object in Table 2 in
Section~\ref{res-sum}, which includes details of angular extent,
morphology and degrees of polarization (see Section~\ref{P}). Nearly
all of our targets have bright, point source-like cores in total flux
and in Table 1 we list the full width half maxima (FWHM) of the
cores. We also list the FWHM of nearby PSF comparison stars to
indicate the seeing conditions at the time of the
observations. Objects are referred to by their {\em
IRAS}~~identifiers.  Position angles (PA) are quoted East of
North. The labels N, E, S, W and combinations thereof are used to
denote compass points.

\subsubsection{A Note on the Degree of Polarization}
\label{P}
In the majority of our targets, the central star is visible in the NIR
and dominates the total flux. Although the polarized flux images
correctly represent the intensity in scattered light from the CSE, the
{\em degree} of polarization will be diluted by the superimposed
unpolarized light from the extended PSF of the star. In order to
calculate the true degree of polarization of the scattered light, the
PSF of the star must be removed to reveal a total flux image of the
nebulosity. Even with a perfect PSF calibration, a PSF subtraction or
deconvolution is likely to be only partially successful due to the
huge dynamic range involved and the compact nature of the scattering
region (i.e. the CSE lies within the wings of the PSF). In practice,
we do not have suitable PSF calibrators for several targets and,
given the non-photometric conditions throughout much of the run, we
have not attempted to correct for the unpolarized light. The degrees
of polarization ($P$) quoted throughout the paper are, therefore,
lower limits on the actual degree of polarization.
Where we quote values for the `maximum polarization' in each object,
this is the maximum degree of polarization detected, using 
$0.43$ arcsec ($3\times3$ pixel) bins, after applying the $2\sigma$ cut
mentioned above. 

\subsection{Individual Objects}

\subsubsection{IRAS 17106-3046}
\label{17106-results}
\begin{figure*}
\epsfxsize=18cm \epsfbox{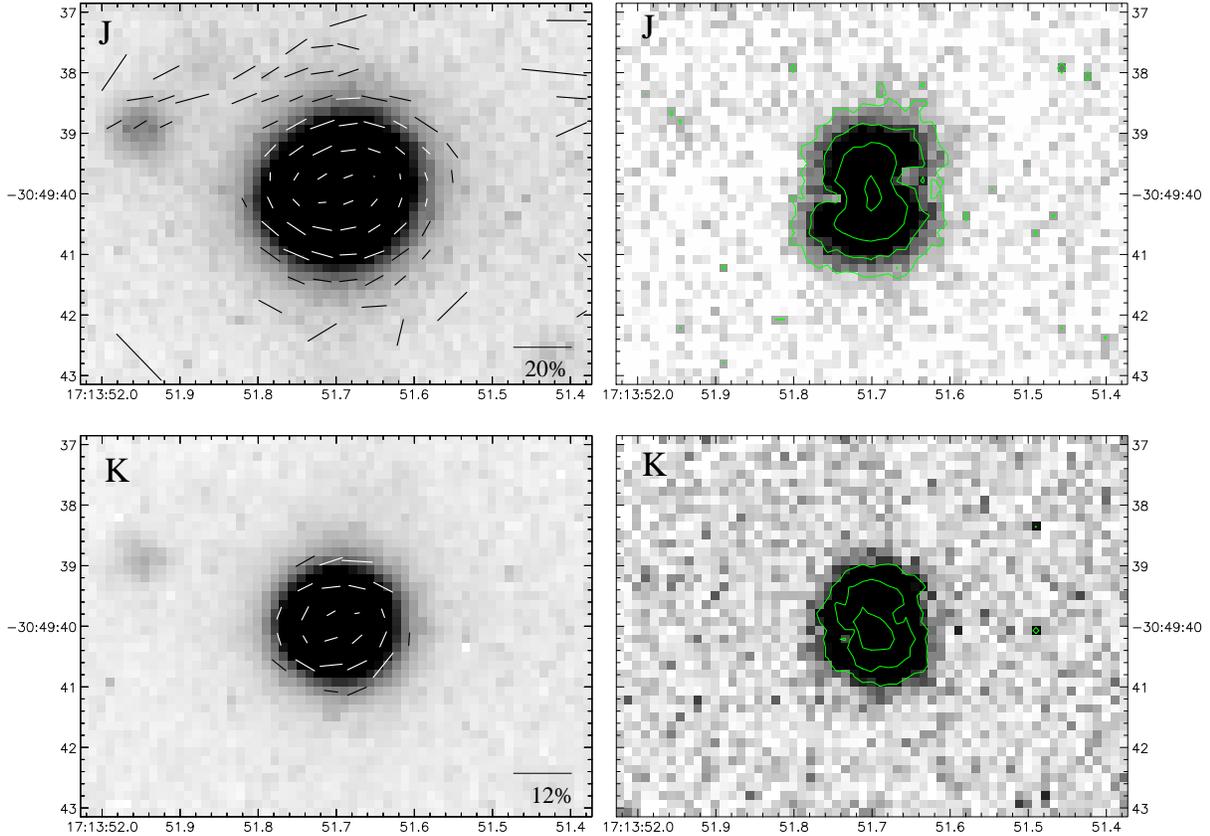}
\caption{Polarimetric observations of IRAS 17106-3046 in the J (upper)
and K (lower) bands. The left hand panels show greyscale images
of total flux superimposed with linear polarization maps (as described
in Section~\ref{P}). The right hand panels show greyscale images of
polarized flux with selected contours superimposed to highlight structure in
the image. The contours are spaced logarithmically at intervals of
1 magnitude. The coordinate system is centred on the peak of the total flux
image, using the coordinates in Table 1,
and is plotted with Right Ascension and Declination (J2000) along the 
horizontal and vertical axes.}
\label{17106-fig}
\end{figure*}

We show J and K band observations of 17106-3046 in
Fig.~\ref{17106-fig}. In the J band the object appears extended and
slightly elongated in the NW-SE direction. Taking a contour level of
five times the RMS fluctuation in the sky background ($5\sigma
_{sky}$) gives an extent of $3.0\times2.6$ arcsec. This is in agreement with
Hrivnak et al. (1999a), who found that 17106-3046 has an angular size
of $2.9$ arcsec in the V band. However, we do not see evidence for
elongation of the bright core in our J and K data, apparent in their V
data. Although the core of the surface brightness image is quite
centrally peaked in both wavebands (Table 1), the J band image shows
evidence for a diffuse halo that extends to the NE of 17106-3046 and
encompasses the faint source found there. This source appears to be a
faint star and may be in association with 17106-3046. The halo is not
apparent in our K band image.

The linear polarization maps shown in Fig.~\ref{17106-fig} demonstrate
that 17106-3046 possesses an extended circumstellar envelope which is
seen by scattered light, forming a reflection nebula. The pattern of
polarization vectors is typical of illumination from a central source,
which is in keeping with our observation that this source is centrally
peaked at infrared wavelengths. However, the pattern is not
centrosymmetric, as would be expected for single scattering in an
optically thin medium with a point source illuminator, and shows
vectors oriented preferentially along a PA of $101$\degr.  This vector
pattern is typical of scattering in an axisymmetric geometry with the
axis at PA 10\degr, perpendicular to the vector orientation.

The polarized flux images trace the regions in which scattering is
occurring and are clearly bipolar. Contours are overlaid on the images
to show detail in the central regions and reveal a bipolar structure
oriented at PA 10\degr, perpendicular to the elongation in the total
flux image. The maximum extent of the object as defined by the
polarized flux is $3.2\times2.4$ arcsec. The bipolar nature of the
polarized flux distribution is further highlighted by the `pinch' in
the contours perpendicular to the bipolar axis in the equatorial
region. At fainter levels, there is evidence for an extension in
polarized flux to the NW of the nebula.  This may indicate a
realignment of the bipolar axis from a PA of 10\degr~~in the central
regions to a more NW direction in the outer regions. In the K band,
the bipolarity in polarized flux is still clear although the nebula is
more compact.  This could be due to our lower detection threshold in
the K band but is also consistent with a scattering cross section that
falls off rapidly with increasing wavelength, as would be typical of
sub-micron dust particles at near-infrared wavelengths.

We also note that there is some evidence for meaningful polarization
in the faint halo to the NE of 17106-3046, extending to the second
faint source. However, the polarization vectors do not conclusively
indicate that this region is illuminated by 17106-3046.

\subsubsection{IRAS 17245-3951}
\label{17245-results}
\begin{figure}
\epsfxsize=11cm \epsfbox[-50 0 347 475]{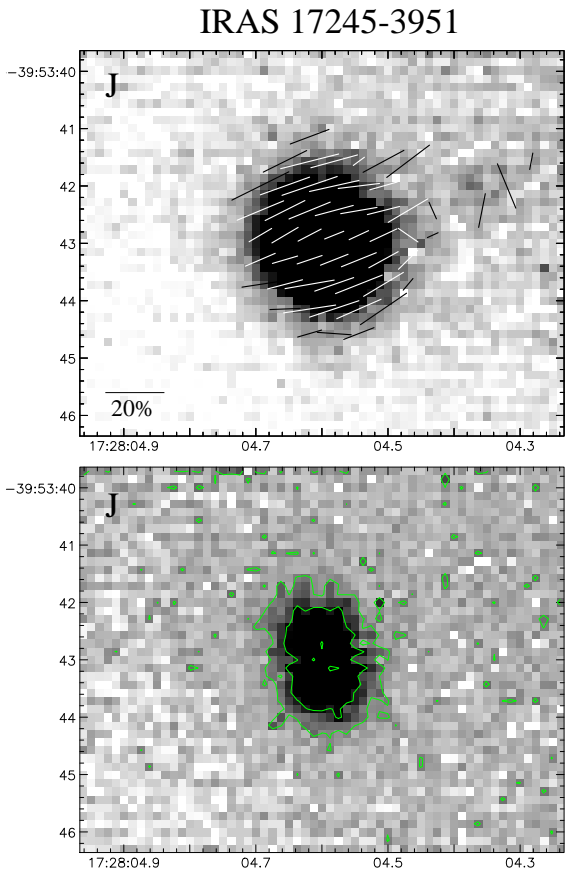}
\caption{Polarimetric observations of IRAS 17245-3951 in the J band. 
The upper panel shows a greyscale image of total flux superimposed with 
a linear polarization map (as described in Section~\ref{P}). The lower 
panel shows a greyscale image of polarized flux with contours superimposed 
to highlight structure in the image. Other details are as in 
Fig.~\ref{17106-fig}.}
\label{17245-fig}
\end{figure}

J band observations of 17245-3951 are shown in
Fig.~\ref{17245-fig}. The greyscale total flux image appears slightly
elongated in a direction just east of north, suggesting that
17245-3951 is a diffuse source at J band and that we are not seeing
through to the star. Taking a contour level at $5\sigma _{sky}$, the
extent of the object in the J band is $3.5\times3.1$ arcsec.

The polarization vectors are aligned at a PA of 103\degr~rather than
being centrosymmetric. Such aligned vector patterns are seen
frequently in observations of embedded Young Stellar Objects (YSOs)
and are typical of scattering in an optically thick and axially
symmetric geometry, such as a dense circumstellar disc or torus
(e.g. Lucas \& Roche 1998). Along this axis away from the centre, the
curvature in the vector pattern indicates that more optically thin
scattering is occurring here, such as would be expected in the lobes
of a bipolar nebula.

The bipolar nature of this object is confirmed by the
polarized flux image shown in Fig.~\ref{17245-fig}. The bipolar axis
is oriented at PA 12\degr, aligned with the elongation in the total
flux and perpendicular to the polarization vectors. This object is
therefore a bipolar reflection nebula illuminated by an embedded
source which is not directly visible in the J band. The extent of the
object in polarized flux is $2.9\times1.6$ arcsec.

Recent WFPC2 imaging by Hrivnak, Kwok \& Su (1999) shows that
17245-3951 is a bipolar nebula at optical wavelengths. They find a
well defined dark lane hiding the source and separating two lobes of
nebulosity. They quote an orientation on the sky of $11$\degr~and an
overall size (including the faint halo) of $2.8\times2.1$
arcsec. These results agree closely with ours suggesting that the
morphology is very similar in the optical and J bands.

\subsubsection{IRAS 17436+5003}
\label{17436-results}
\begin{figure*}
\epsfxsize=22cm \epsfbox[50 0 677 484]{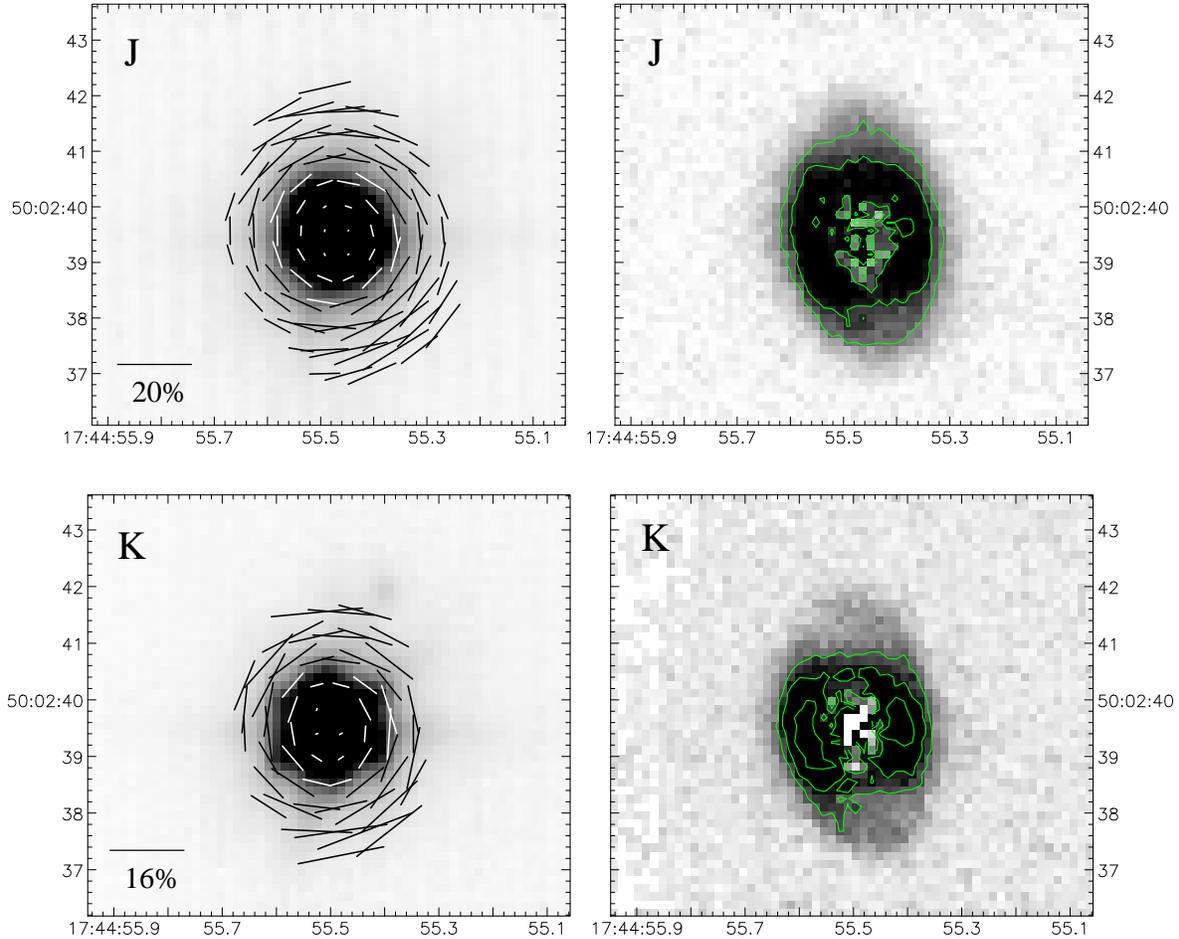}
\caption{Polarimetric observations of IRAS 17436+5003 in the J and K 
bands. Details as for Fig.~\ref{17106-fig}.}
\label{17436-fig}
\end{figure*}

In Fig.~\ref{17436-fig} we show J and K band observations of
17436+5003. The total flux images show no evidence for extended
emission suggesting that the star is directly seen.  Although the
direct images show only a point-like structure, the linear polarimetry
reveals that 17436+5003 does have an extended envelope and that this is
seen by scattered light. High degrees of polarization are evident in
both J and K bands forming an almost perfect centrosymmetric pattern,
indicating scattering in an optically thin medium with a point source
illuminator. The similarity of the polarimetry in the J and K bands
(Table 2) implies that (i) small dust grains ($<0.1\mu$m) are
responsible for scattering the light and (ii) the scattering geometry
is very similar at J and K (i.e. there are no optical depth
effects). The polarization pattern extends more in the N-S direction
than in the E-W direction, suggesting that the reflection nebula is
slightly elongated. In addition, the polarizations in the N-S
direction are larger by up to a factor of 2 than those in the E-W
direction which could be interpreted as a narrowing in the range of
scattering angles (along the line of sight) in the N-S
direction. These subtle deviations from centrosymmetry imply that the
scattering geometry is not spherically symmetric and instead possesses
an element of axisymmetry. The degrees of linear polarization towards
the core of 17436+5003 decrease to very small values. These regions
are heavily affected by the bright PSF from the star so that any light
polarized by scattering in the stellar envelope will be greatly
diluted by the unpolarized light from the star itself.
\begin{figure*}
\epsfxsize=18cm \epsfbox{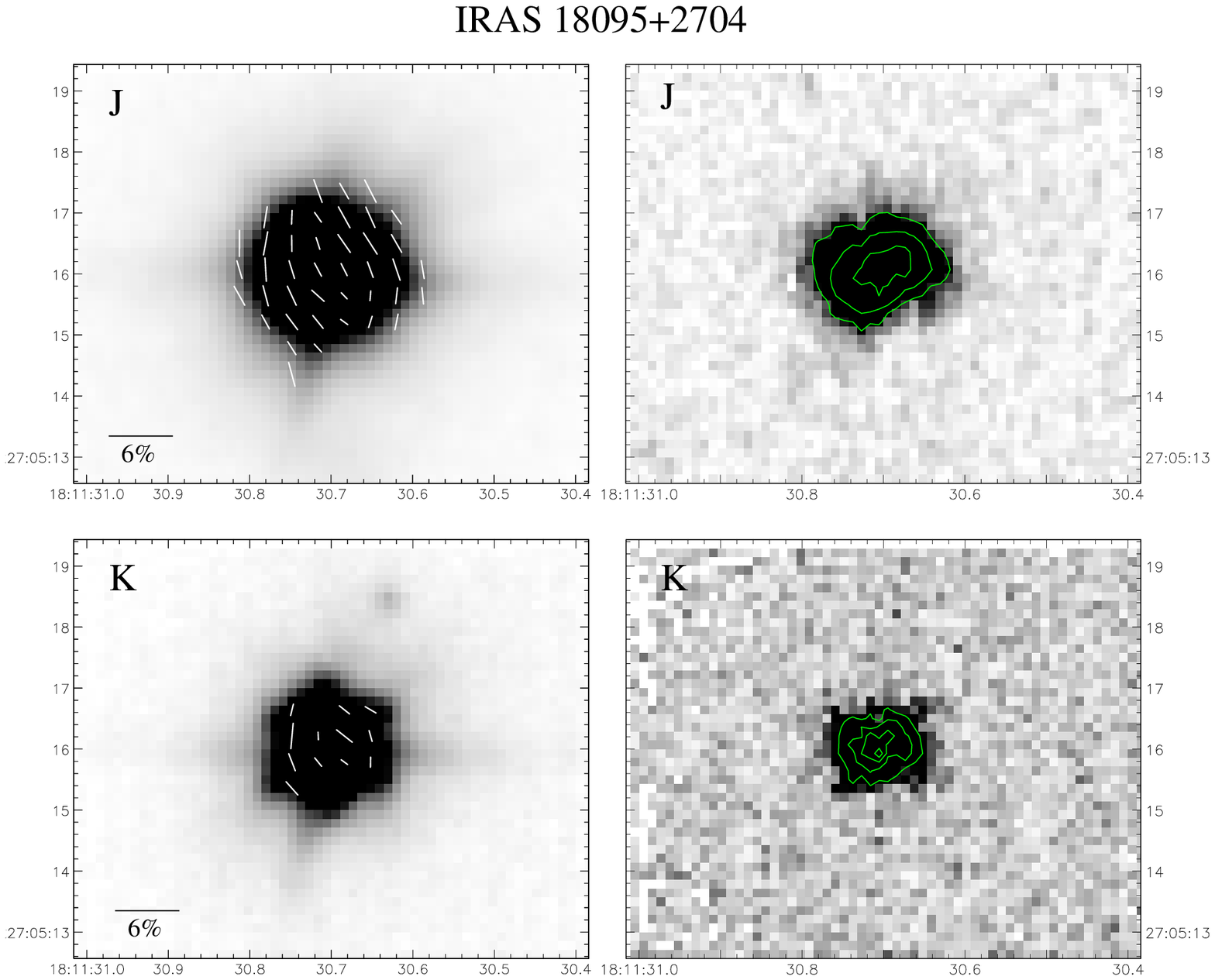}
\caption{Polarimetric observations of IRAS 18095+2704 in the J and K
bands. Details as for Fig.~\ref{17106-fig}.}
\label{18095-fig}
\end{figure*}
The polarized flux images shown in Fig.~\ref{17436-fig} bear no
resemblence to the direct images and reveal the true nature of the
scattering envelope surrounding this star. Most scatterring occurs in
two bright arcs on either side of the source with a fainter elliptical
halo of scattered light extending to the N and S. The core of the
polarized flux image is contaminated by residuals from the bright
stellar PSF but sufficient information remains to clearly reveal the
shell-like structure of the scattered light.

We suggest a geometry for 17436+5003 consisting of an ellipsoidal
dusty envelope surrounding the star. Scattering of light within this
envelope results in the faint halo seen in polarized flux. The bright
arcs then result from limb-brightening as light is scattered within a
shell-like density enhancement embedded within the envelope. The
thickness of the shell is difficult to assess quantitatively but a cut
through the arcs in an E-W direction gives a FWHM of $0.86$ arcsec in
the J band. In the K band the arcs appear shorter and thicker. The
shell is thinner in the polar regions, along the major axis at PA
10\degr~~suggesting an axisymmetric density structure which could act
to collimate an outflow along this axis.

In HST/WFPC2 imaging observations (Ueta et al. 2000) 17436+5003
appears as an elliptical nebula at optical wavelengths, with a similar
extent and orientation to that shown in our NIR data.  The object was
just resolved in $10.5 \mu$m and $12.5 \mu$m observations by Skinner
et al. 1994 whose deconvolution revealed a dusty torus oriented
perpendicular to the major axis of the optical (Ueta et al 2000) and
NIR (this paper) emission. They estimated an inner radius of $0.48$
arcsec for this torus. This agrees remarkably well with the
$0.6\pm0.2$ arcsec inner radius of the shell seen in the polarized
flux images of Fig.~\ref{17436-fig}. It seems highly likely therefore,
that the dust `torus' seen in emission by Skinner et al. (1994) is
also responsible for scattering much of the light in our J and K band
data and represents an optically thin, ellipsoidal shell of dust with
an equatorial density enhancement. An optically thin nature is
required to explain the lack of wavelength effects between J and K
band in the polarization data and to explain the fact that the optical
(WFPC2) data is morphologically very similar to our NIR data. The
sharpness of the inner edge of the shell in Fig.~\ref{17436-fig}
indicates that it is physically detached from the star and that the
interior is relatively empty of dust. This is consistent with the
Class IVb SED of 17436+5003 (Ueta et al. 2000; van der Veen, Habing \&
Geballe 1989).

\subsubsection{IRAS 18095+2704}
\label{18095-results}

Observations of 18095+2704 are presented in Fig.~\ref{18095-fig}. In
both J and K total flux the object appears as a bright point source
with the telescope diffraction spikes visible.  The polarization maps
indicate that scattering is indeed occurring and that 18095+2704 has
an extended envelope forming a reflection nebula at NIR
wavelengths. In the J band the polarization vectors appear aligned
preferentially along an axis at PA 23\degr~~rather than in a
centrosymmetic arrangement, which is indicative of multiple scattering
in an axisymmetric or bipolar geometry. A similar pattern is seen at K
although the low signal to noise in the scattered component precludes
a detailed analysis of the pattern. The low polarization in this
object (Table 2) is probably mostly due to the diluting effect of the
unpolarized light from the bright star itself, however, the degree of
polarization in the wings of the PSF is still only $2.4$ per cent
maximum.  This compares with polarizations of $20$ per cent in
17436+5003, a similarly bright and point source-like object, which
suggests that the scattering region in 18095+2704 is more compact than
in 17436+5003 and therefore more strongly influenced by the stellar
PSF. In addition, if a bipolar geometry is advocated then the low
degrees of polarization could be due to a high inclination of the
bipolar axis to the plane of the sky.

The polarized flux images shown in Fig.~\ref{18095-fig} reveal a
bipolar geometry for the scattered light in 18095+2704. The
orientation of the bipolar axis is at PA 115\degr, perpendicular to
the preferential orientation of the polarization vectors. In both
wavebands the brightest point of the polarized flux image is offset to
the SE relative to the peak of the total flux image, which gives the
position of the star, so that the SE lobe of the bipolar is larger
than the NW lobe. This also suggests that the bipolar axis is inclined
to the plane of the sky and that the SE lobe is oriented towards us.

This interpretation is supported by the WFPC2 V and I band images of
Ueta et al. (2000) which show a bipolar nebula oriented at a similar
position angle to that in our polarized flux images and in which the
SE lobe appears the larger and brighter of the two. These authors also
comment on the brightness of the central star and in their
classification of this object suggest that the star is seen
directly. This constrasts with the clear bipolar nature of the object
in scattered light, indicating that there must be a dusty disc or
torus surrounding the star that is responsible for collimating the
outflow into two bipolar lobes. The aligned vectors in the
polarization patterns at J and K indicate that dust grains in the
bipolar lobes do not have a direct view of the source and that
multiple scattering is important. If the optical depth within the
scattering region is greater than one in the NIR then it seems
unlikely that we are seeing the star directly in the V and I
bands. However, the star does appear very bright and point
source-like, although our J band data show evidence for elongation in
the core along the bipolar axis. If this axis is highly inclined to
the plane of the sky then it is possible that we have a preferential
view of the star and that our line of sight lies within the
illumination cone of the SE lobe. Trammell, Dinerstein \& Goodrich
(1994) measure a polarization of $4.5$ per cent at 750 nm rising to
$8.5$ per cent at 480 nm for this object, which also indicates that
the star is more obscured at shorter wavelengths.

\subsubsection{IRAS 18184-1623}
\label{18184-results}
\begin{figure}
\epsfxsize=11cm \epsfbox[30 0 427 474]{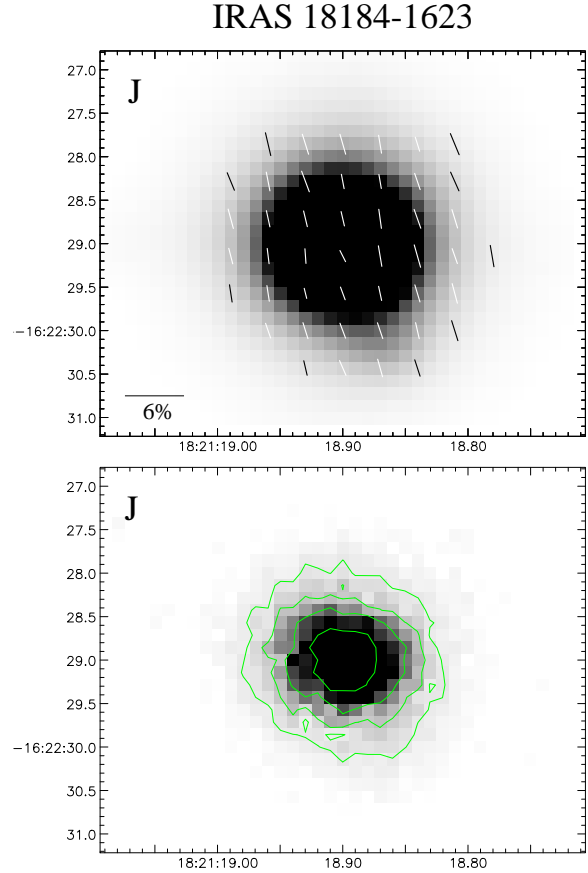}
\caption{Polarimetric observations of IRAS 18184-1623 in the J band. 
Details are as for Fig.~\ref{17245-fig}.}
\label{18184-fig}
\end{figure}

Observations of 18184-1623 in the J band are shown in
Fig.~\ref{18184-fig}. There is evidence for an elongation in the core
to the SW of the stellar centroid. Due to the brightness of this
target and the fast readout mode (Table 1) it is possible that
spurious effects may have been introduced. However, observations of a
PSF standard star, taken just after 18184-1623 in the same mode, have
a similar FWHM to the target object but show no evidence of
elongation. We conclude that the elongation in the core is therefore
real.

The polarization pattern is also unusual, consisting of almost
parallel vectors oriented at a PA of approximately $20$\degr. Since a
total of 72 exposures of 18184-1623 were taken, we examined subsets of
the data for consistency. The polarization pattern was found to be
robust within our data set and not due to a few anomalous exposures.
The parallel orientation of the vectors would normally signify
scattering in a very optically thick geometry, such as a dense dusty
torus around the star. However, such an axisymmetric scattering
scenario would be expected to produce a bipolar geometry in polarized
flux which is not seen. The polarized flux image shown in
Figure~\ref{18184-fig} is point-like and coincident with the peak in
total flux. It is possible that we simply have not resolved the
bipolar structure. We note, though, that there is slight evidence for
an extension of the polarized flux contours to the SW matching the
previously noted extension in the total flux.

At MIR wavelengths most of the flux comes from an extended ($12\times16$
arcsec) elliptical nebula with two well defined peaks
(Robberto \& Herbst 1998). These authors suggest that the emission
results from `a relatively thin edge-brightened shell'. A similar
morphology is seen by Meixner et al. (1999). Interestingly, the axis
of this elliptical dust shell lies at PA 114\degr (measured from
Robberto \& Herbst's $10.1 \mu$m image) which is approximately
perpendicular to the vector orientation in our polarization map. If
the parallel orientation of the vectors is indicative of scattering in
an optically thick dust torus then the polar axis of the torus would
lie parallel to the major axis of the extended MIR nebula.

\subsubsection{IRAS 19114+0002}
\label{19114-results}
\begin{figure*}
\epsfxsize=18cm \epsfbox{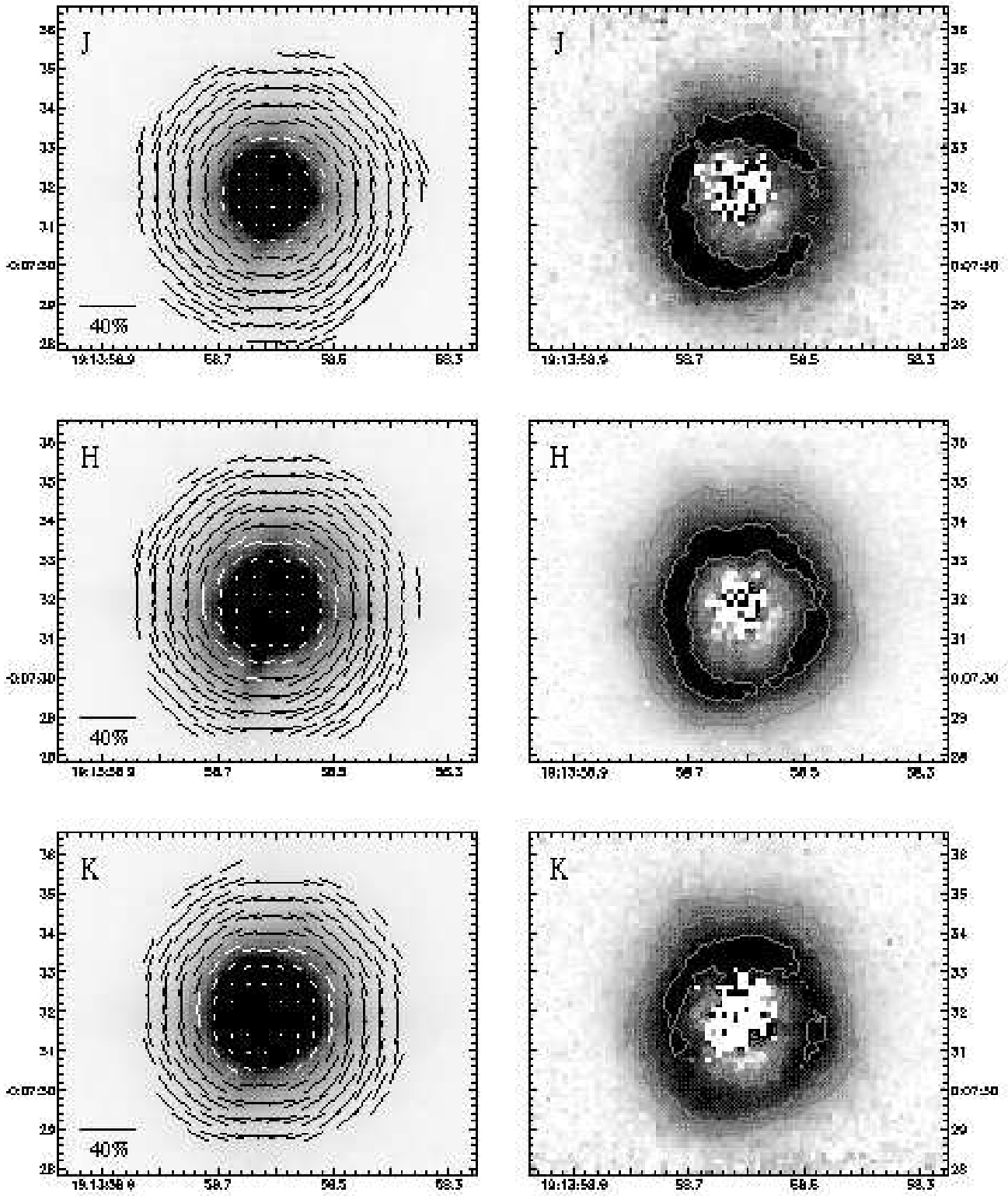}
\caption{Polarimetric observations of IRAS 19114+0002 in the J, H and
K bands. The polarized flux contours are spaced logarithmically at
intervals of $0.5$ magnitudes. Other details are as for
Fig.~\ref{17106-fig} with the addition of the H band observations.}
\label{19114-fig}
\end{figure*}

Observations of 19114+0002 in J, H and K bands are shown in
Fig.~\ref{19114-fig}. The bright, point source-like nature in total
flux is evident in all three wavebands (see Table 1).  The extent of
the image in total flux (defined by a contour at $5\sigma _{sky}$) is
50 pixels ($7.2$ arcsec) in all three bands.

The polarization maps shown in Fig.~\ref{19114-fig} show that the star
illuminates an extensive reflection nebula. Polarization is reliably
detected out to the full extent of the total flux image. The vector
pattern is centrosymmetric to a high degree in all three wavebands,
which suggests that the nebula is illuminated isotropically; on the
basis of the polarization pattern there is no obvious evidence for
axisymmetry.  Although the direct light from the star reduces the
degrees of polarization in the core region to very low values, the
polarizations in the extended nebula are high (Table 2).  These values
will still be contaminated to some degree by unpolarized light in the
wings of the stellar PSF and to recover the true degree of
polarization and its radial dependence the PSF will need to be removed
(effectively forming the total flux image of the scattering
nebula). However, the similarity of degree of polarization in the
three infrared bands suggests that there are no strong wavelength
effects at play as far as scattering is concerned and that the
envelope is, therefore, optically thin.

The polarized flux images show that most of the scattered light
originates from a ring of dust surrounding the star. The width of the
ring, determined by taking a radial profile in polarized flux and
estimating the FWHM, varies between $0.8$ and $1.3$ arcsec in all
three bands. The inner radius varies between $1.1$ and $1.6$ arcsec
and the outer radius extends to $2.3$ arcsec. This brighter structure
is then embedded within a more extensive and fainter halo of scattered
light which is traced out to a radius of $\sim4.5$ arcsec in our
images. The central regions of our polarized flux image are
contaminated by residuals due to errors in the registration,
normalization and subtraction of the bright core-dominated images,
although steps have been taken during the reduction stages to minimize
this. However, sufficient information is recorded interior to the
bright ring to reveal that polarized flux above the background, and
hence dust, is present in this region and that it is scattering light.
In particular, the polarization vector patterns extend inwards beyond
the inner boundary of the bright ring and retain their centrosymmetric
structure.

The morphology of the ring, as revealed by the polarized flux
distribution, is not entirely symmetric and there are subtle
differences between the three wavebands. In particular, the ring
thickness varies with azimuthal angle and appears thinnest to the SW
at a PA of approximately 195\degr.  This `thinning' increases with
wavelength until in the K band it becomes almost a gap in the
ring. In the K band the ring structure appears more clumpy with the
northern portion of the ring the brightest. The stellar centroid is
not located at the centre of the ring but is instead displaced about 2
pixels ($\sim0.3$ arcsec) to the NE. This displacement was also noted
by Jura \& Werner (1999) in their MIR images.

\begin{figure*}
\epsfxsize=18cm \epsfbox{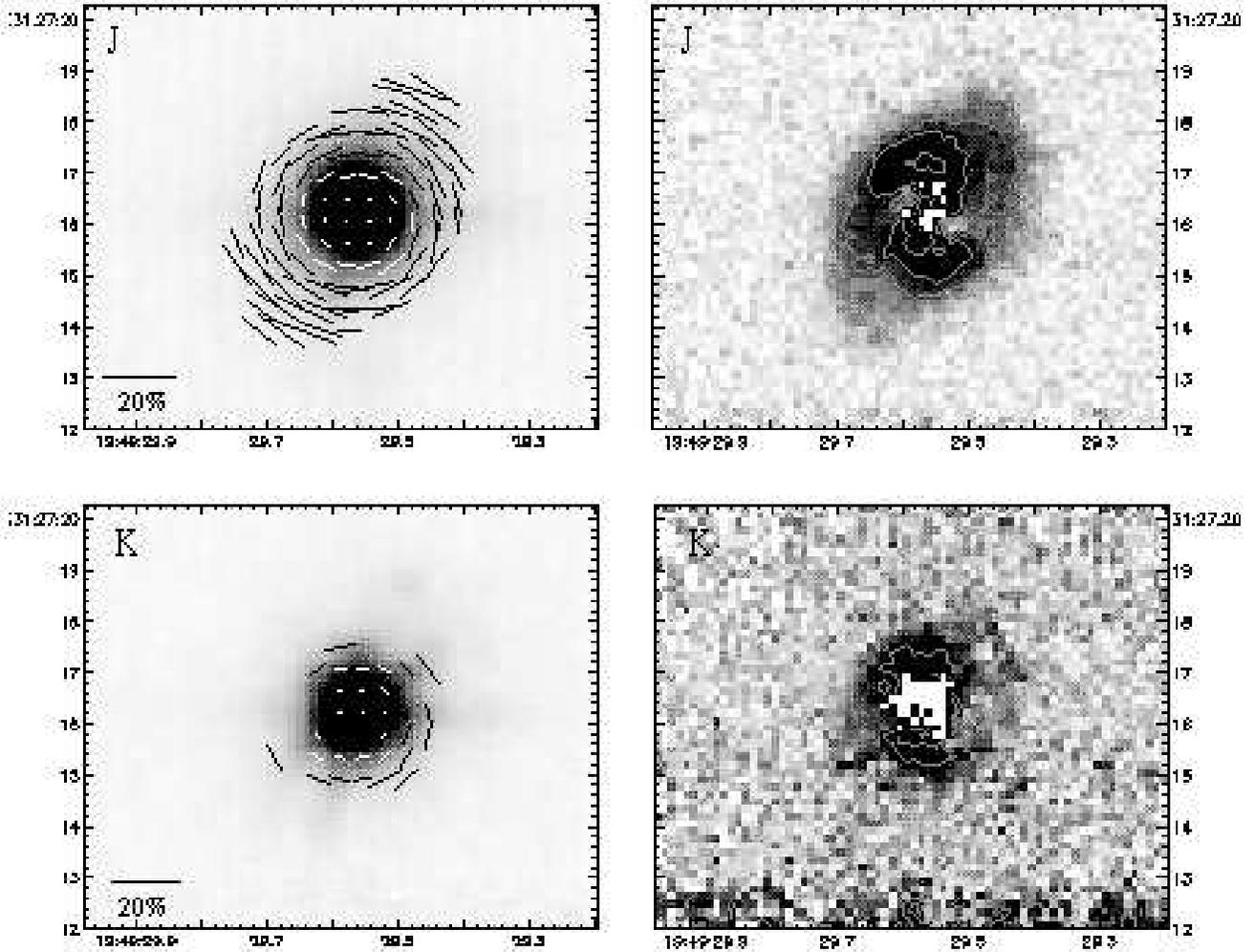}
\caption{Polarimetric observations of IRAS 19475+3119 in the J and K
bands. The polarized flux contours are spaced logarithmically at
intervals of $0.5$ magnitudes. Other details are as for Fig.~\ref{17106-fig}.}
\label{19475-fig}
\end{figure*}

Coronographic imaging polarimetry of 19114+0002 was obtained by
Kastner \& Weintraub (1995) who detected the extended scattering
nebula out to a radius of $9$ arcsec at J and $6$ arcsec at K. This
suggests that the dust halo is at least twice as large as shown in our
non-coronographic J band images. At a radius of $5$ arcsec they
measure polarizations of $13$ percent at J and $6$ per cent at K. A
radius of $5$ arcsec corresponds to the outermost part of the envelope
in our detection, where the polarizations are largest.  Here we
measure polarizations of greater than $25$ per cent in both J and
K. We cannot account for the discrepancy since any correction for the
underlying stellar PSF in our data would only serve to increase the
degree of polarization. Hawkins et al. (1995) present MIR imaging of
19114+0002 in which their deconvolved $12.5 \mu$m image shows a dust
ring encircling the star. This ring, seen in thermal emission, is
remarkably similar in both size and morphology to the ring of
scattered flux seen in our data and they undoubtedly arise from the
same physical structure. The image of Hawkins et al. (1995) also shows
the `thinning' of the ring to the SW of the star seen in our data. In
the WFPC2 survey of Ueta et al. (2000), 19114+0002 appears as an
extended nebula with a complex structure of concentric shells forming
a rosette. The extent of the nebula in these optical images is similar
to that of out NIR images, being approximately $8$ arcsec in diameter.
They also show clear evidence for a bipolar `protruberance' at PA
20\degr~~which appears to extend beyond the boundary of the
nebula. The SW part of this feature is coincident with the thin part
of the dust ring seen in our data and in the MIR image of Hawkins et
al. (1995) and may be due to a more recent and more highly collimated
outflow breaking out through the dust ring. This is also consistent
with our observation that the ring is extended to the SW so that the
star no longer resides at its centre.

The high quality of our polarimetric results in three infrared wavebands
will allow us to construct detailed scattering models to analyse the
structure of the circumstellar envelope in 19114+0002 and these will be
presented in a subsequent paper.

\subsubsection{IRAS 19454+2920}
\label{19454-results}

This is the only source in our sample which shows no evidence of
polarization. It has been classified as a PPN on the basis of its
infrared colours. The {\em IRAS} flux peaks in the $25 \mu$m band
which indicates a dust temperature of $188$ K (Volk \& Kwok 1989) and
its position in the {\em IRAS} colour-colour diagram is similar to
other objects in our sample (e.g. 19500-1709). Our polarized flux
images (not shown) show only noise and residuals from image
alignment. If the envelope is smaller than $\sim0.5$ arcsec on the sky
then we will not have resolved it. However, if axisymmetry is present in
the envelope than even in an unresolved observation we would expect to
observe a net polarization (equivalent to aperture polarimetry). We
conclude therefore that the CSE in this case must have an angular size
$<0.5$ arcsec with no significant axisymmetry (i.e. it is optically
thin in the NIR and/or illuminated isotropically).

\subsubsection{IRAS 19475+3119}
\label{19475-results}

Our J and K band observations of 19475+3119 are shown in
Fig.~\ref{19475-fig}. In total flux the object appears as a bright
star with the diameter of the stellar profile (defined using a contour
at $5\sigma_{sky}$) being $\sim5$ arcsec at J. The intensity contours are
similar to those of the PSF star and show no evidence for deviation
from sphericity.

The polarization maps shown in Fig.~\ref{19475-fig} show that
19475+3119 does possess an extended scattering envelope. The
polarization pattern at J is centrosymmetric indicating that the star
is embedded within and isotropically illuminates a reflection
nebula. The polarizations fall to very low values in the core
suggesting that the star is seen directly and depolarizes the
scattered light. The region of scattering is more extended in the
SE-NW direction at a PA of 145\degr~~so that the envelope is not
spherical. In the K band the nebula appears much less extensive
suggesting a lower surface brightness at longer wavelengths, which
would be consistent with scattering from dust grains much smaller than
the wavelength (Rayleigh scattering, which results in a $\lambda^{-4}$
fall off in scattering cross section with increasing
wavelength). Nevertheless, a centrosymmetric polarization pattern is
still apparent at K.

In polarized flux the scattering nebula is clearly elongated in the J
band data. The fainter emission forms a halo with dimensions of
$4.9\times3.4$ arcsec with the major axis oriented at PA 145\degr.
Superimposed upon this is a remarkable structure which resembles two
spiral arms arranged symmetrically about the star. Since the
centrosymmetry of the polarization pattern indicates that the
illumination of the nebula is isotropic, these structures must
represent concentrations of optically thin dust. The dust
concentrations are neither parallel nor perpendicular to the long axis
of the nebula but instead have a rotational point symmetry about the
star. In the K band, the core of the polarized flux image is
contaminated by residuals from the bright stellar PSF and the faint
halo is only barely detected.  However, the peaks in polarized flux on
either side of the star, seen in the J band data, are also seen at K.

Previous observations of 19475+3119 have failed to detect evidence of
an extended envelope around this source. In their optical imaging
observations, Hrivnak et al. (1999a) concluded that there was no
evidence for an extended nature. This is not surprising given the
faintness of the scattering halo, the compact nature of the nebula and
that the brightest regions of scattered light are distributed
symmetrically about the star. These observations illustrate that high
spatial resolution imaging polarimetry is a very effective technique
for detecting and imaging extended scattering envelopes around bright 
stars. This object is discussed further in Section~\ref{spiral}.

\subsubsection{IRAS 19477+2401}
\label{19477-results}
\begin{figure}
\epsfxsize=11cm \epsfbox[-10 0 387 475]{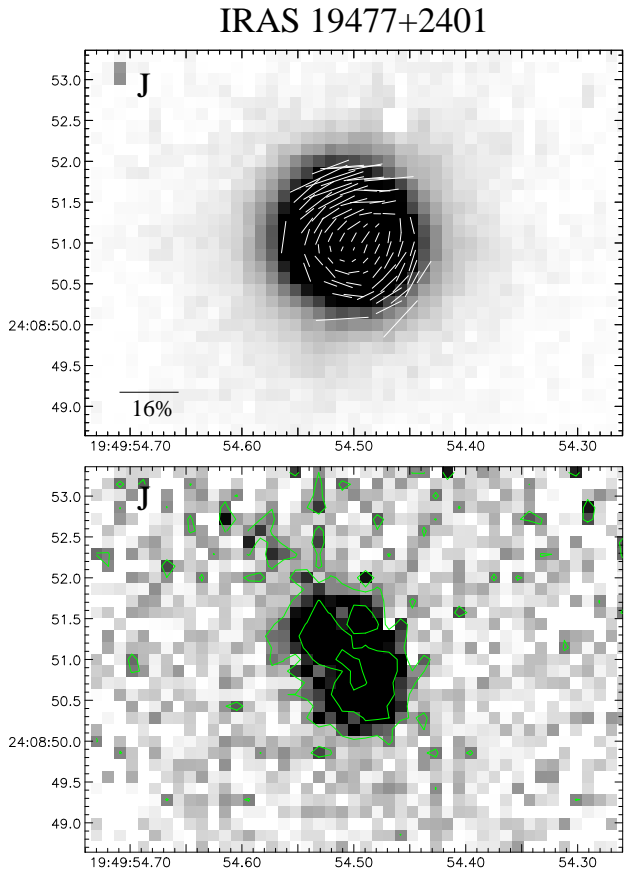}
\caption{Polarimetric observations of IRAS 19477+2401 in the J band. 
Details are as for Fig.~\ref{17245-fig}.}
\label{19477-fig}
\end{figure}

J band observations of 19477+2401 are shown in Fig.~\ref{19477-fig}.
The imaging polarimetry data shown in Fig.~\ref{19477-fig} shows that
19477+2401 is clearly extended and is seen in scattered light. Due to
the compact nature of this object we have chosen to display the
polarization vectors without any spatial binning so that each vector
represents a measurement in a single pixel. Since the pixels are
smaller than the seeing size ($0.143$ arcsec compared with a typical
seeing disc of $0.5$ arcsec) the measurements are not strictly
independent. However, a polarization pattern indicative of scattering
is clearly seen with vectors away from the core having a distinct
centrosymmetric arrangement. In the core region vectors are aligned at
PA $\sim125\degr$ suggesting multiple scattering in an optically thick
environment, such as a dusty circumstellar disc.

In J band polarized flux, 19447+2401 appears as a bipolar reflection
nebula. Two lobes of nebulosity are visible, separated by a dark lane.
The SW lobe appears to be the brighter of the two and is also the more
highly polarized. This suggests that the bipolar axis is inclined to
the plane of the sky and that the SW lobe is oriented towards
us. Hrivnak et al. (1999a) noted that 19477+2401 was very faint in
their V band data and was `extremely red'. This would agree with our
interpretation of a star surrounded by an optically thick disc which
channels the illumination into two lobes to form a bipolar nebula. Our
polarimetry observations suggest that light from the core of this
object is mutiply scattered so that, despite the stellar appearance in
total flux, the star is not seen directly even in the J band.

\subsubsection{IRAS 19500-1709}
\label{19500-results}
\begin{figure*}
\epsfxsize=16cm \epsfbox{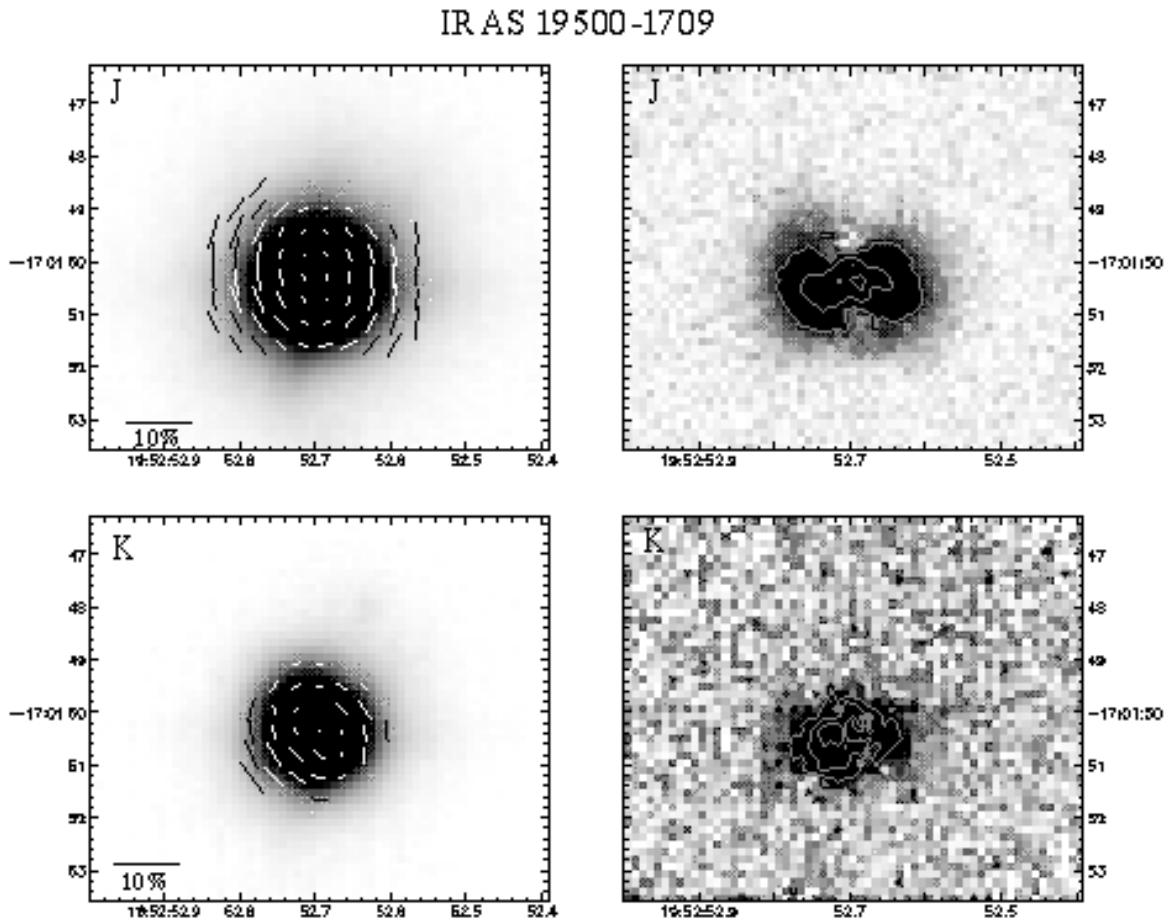}
\caption{Polarimetric observations of IRAS 19500-1709 in the J and K
bands. The polarized flux contours are spaced logarithmically at
intervals of $0.8$ magnitudes. Other details are as for Fig.~\ref{17106-fig}.}
\label{19500-fig}
\end{figure*}

J and K band observations of 19500-1709 are shown in
Fig.~\ref{19500-fig}.  In total flux there is no obvious indication of
extension in this object beyond that of the PSF or of deviation in the
isophotes from those expected of a point source.  The polarization
maps show that the star illuminates a reflection nebula which extends
to a radius of $>2$ arcsec in the J band. Although the polarization
pattern appears approximately centrosymmetric, there are deviations
which indicate that the illumination is not isotropic. In particular,
the polarizations are higher along an E-W axis than along a N-S
axis. In addition, there is evidence that the vectors are oriented
preferentially N-S in the core region, suggesting that multiple
scattering is occurring in this region. The J band polarized flux
image clearly shows that the illumination is indeed anisotropic and
that 19500-1709 is a bipolar nebula in scattered light. The major axis
of the bipolar, as defined by the brighter regions of polarized flux,
is oriented at PA $100\degr$. The two lobes are separated by a dark
lane oriented perpendicular to the major axis.  This elongated bright
structure is superimposed upon a fainter halo which, although less
collimated in appearance, is still evidently bipolar. The major axis
of the faint halo is oriented more E-W than that of the bright
structure, however, suggesting that this object has two axes separated
by approximately $10\degr$.

In the K band, a similar polarization pattern is seen, although the
nebula appears more compact than in the J band data. The K band
polarization map also shows evidence for anisotropic illumination and
this is confirmed by the polarized flux image which, although noisier
than the J band image, is identifiably bipolar. Interestingly, the
major axis of the nebula in polarized flux in the K band appears
rotated by approximately $10\degr$ relative to that at J. In both the
J and K polarized flux images the E lobe appears brighter suggesting
that the bipolar axis is tilted out of the plane of the sky so that
this lobe points towards us.

The existence of an outflow from 19500-1709 is suggested by the
detection of a fast ($>40$ kms$^{-1}$) velocity component in the
$^{12}$CO $J=2-1$ and $J=1-0$ lines (Bujarrabal, Alcolea \& Planesas
1992). However the outflow is not resolved in their data nor in the
$^{12}$CO $J=2-1$ data of Neri et al. (1998). In optical
spectropolarimetry observations, Trammell et al. (1994) did not detect
instrinsic polarization in 19500-1709 within their aperture. Our
imaging polarimetry observations clearly show that this object is
intrinsically polarized and appears as a bipolar reflection nebula in
the NIR.

\subsubsection{IRAS 20000+3239}
\label{20000-results}
\begin{figure*}
\epsfxsize=18cm \epsfbox{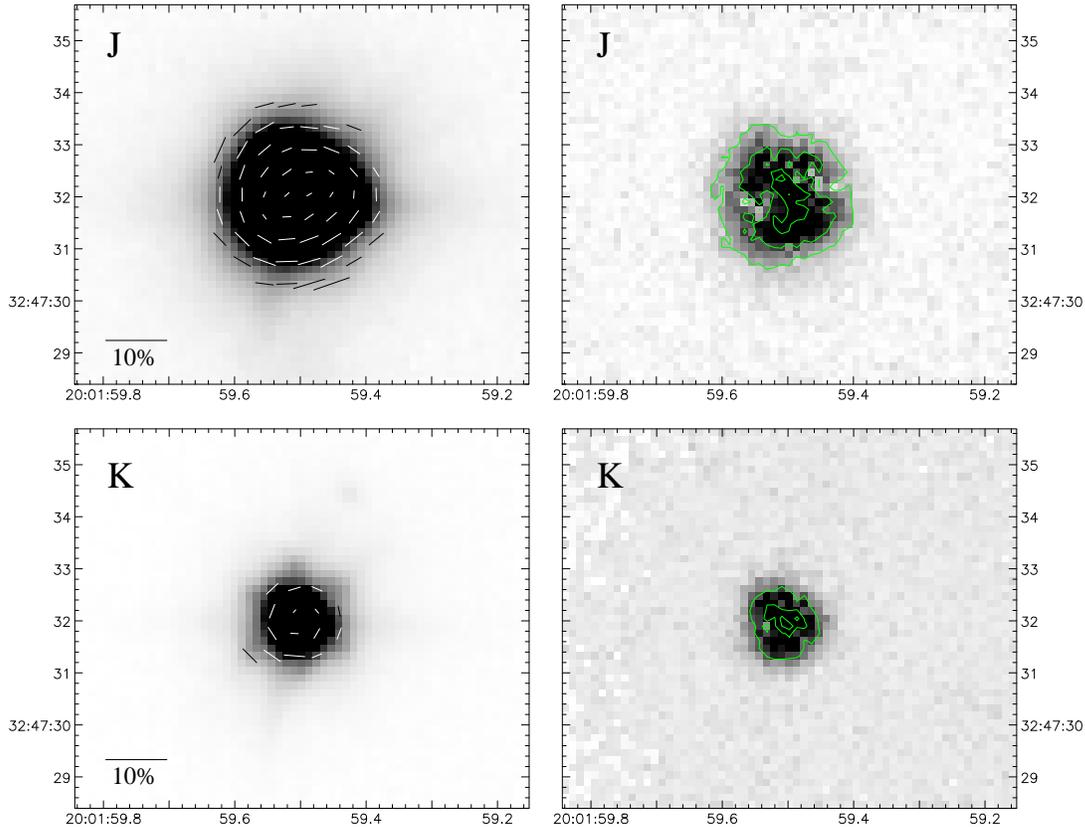}
\caption{Polarimetric observations of IRAS 20000+3239 in the J and K
bands. Details are as for Fig.~\ref{17106-fig}. }
\label{20000-fig}
\end{figure*}

Observations of 20000+3239 are shown in Fig.~\ref{20000-fig} for J and
K bands. The polarization map clearly shows that 20000+3239 has an
extended scattering envelope that is illuminated by the central
star. The extent of the recorded polarization pattern in the J band is
$3.5$ arcsec in diameter. The vector pattern is approximately
centrosymmetric in the outer regions of the nebula but shows an
increasing preference for vector alignment along a PA of $120\degr$
closer to the core. The core region is significantly polarized, even
with the underlying stellar flux unsubtracted. This suggests that
multiple scattering is occurring in the core region and is consistent
with a geometry in which the star is surrounded by an obscuring disc
or torus. This then results in anisotropic illumination of the outer
regions and the deviation from centrosymmetry in the polarization
pattern in these regions. In this interpretation the equatorial plane
of the disc/torus would also be at PA $120\degr$. In the K band,
scattered light is detected but from a less extended area. The
polarization pattern also appears more centrosymmetric than at J, with
less evidence for aligned vectors in the core region.  This is
consistent with scattering in an optically thick dust distribution
surrounding the star, which becomes progressively more transparent
with increasing wavelength, producing a more centrosymmetric
polarization pattern at the longer wavelengths.
\begin{figure*}
\epsfxsize=18cm \epsfbox{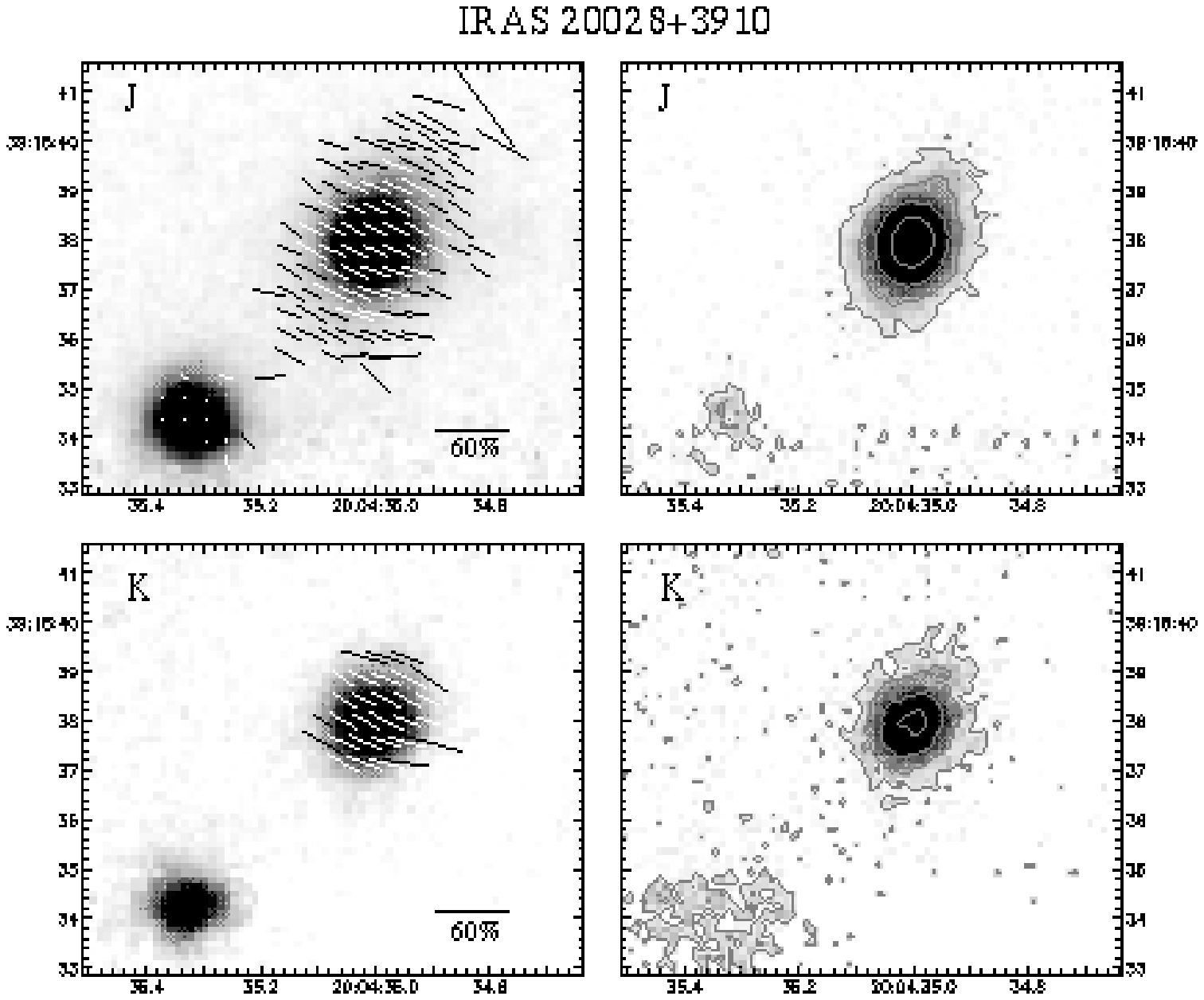}
\caption{Polarimetric observations of IRAS 20028+3910 in the J and K
bands. Details are as for Fig.~\ref{17106-fig}.}
\label{20028-fig}
\end{figure*}
In polarized flux the scattered light distribution is bipolar. This is
particularly evident in the J band, where a bipolar structure is seen
superimposed upon a fainter and more elliptical halo. The PA of the
major axis is at $36\degr$, approximately orthogonal to the aligned
vector orientation and the postulated circumstellar disc plane. The
bipolarity is accentuated by the equatorial `pinching' in the
polarized flux image which is characteristic of scattering in an
axisymmetric geometry. The extent of the faint halo (to a level of
$5\sigma_{sky}$ in polarized flux) is $3.7$ arcsec along the major
axis and $3.3$ arcsec perpendicular to it. In the K band the bipolar
structure is still discernible but obviously more compact. The halo
extent in this case is $2.2\times1.9$ arcsec.

This object was observed by Hrivnak et al. (1999a) who, on the basis of
their ground-based optical imaging, concluded that it is `slightly
extended' and `round in shape'. In the MIR imaging survey of Meixner
et al. (1999), 20000+3239 appeared `marginally extended' with a
possible E-W elongation.  Our polarimetry data clearly shows that this
object is both extended and axisymmetric with the star illuminating a
bipolar reflection nebula.  Given the evidence for multiple scattering
in the core region, it is unlikely that the star is seen directly at
optical or NIR wavelengths despite its brightness. 

\subsubsection{IRAS 20028+3910}
\label{20028-results}

Observations of 20028+3910 in the J and K bands are shown in
Fig.~\ref{20028-fig}. The object appears extended and elongated in
total flux in the J band, especially when compared to the stellar
source in the bottom left of the image. The extent of the faint
emission, using an isophote at $5\sigma_{sky}$, is $4.0\times3.3$
arcsec with the major axis at PA $153\degr$. At the highest flux
levels, the core is also extended in the same direction so that
20028+3910 does not have a stellar profile in the NIR.

The polarization maps show a pattern of aligned vectors over most of
the object. In both wavebands the vectors in the core region are
aligned at PA $65\degr$, orthogonal to the major axis of the
object. In the J band, there is some evidence for a scattering pattern
in the outer regions (with vectors deviating slightly from the aligned
pattern) but the polarization map is on the whole typical of
scattering in an optically thick environment. The aligned vectors over
the source indicate that little or no direct light from the star
reaches us and that the light that we do see is multiply
scattered. This suggests a geometry in which the star is surrounded by
an optically thick circumstellar torus with equatorial plane at PA
$65\degr$ which beams the illumination along its polar axis to form a
bipolar reflection nebula. The lack of a centrosymmetric polarization
pattern also indicates that dust grains within the nebula do not see
the source directly (except perhaps on the bipolar axis). In the K
band we have only detected polarization in the core region.
\begin{figure*}
\epsfxsize=18cm \epsfbox{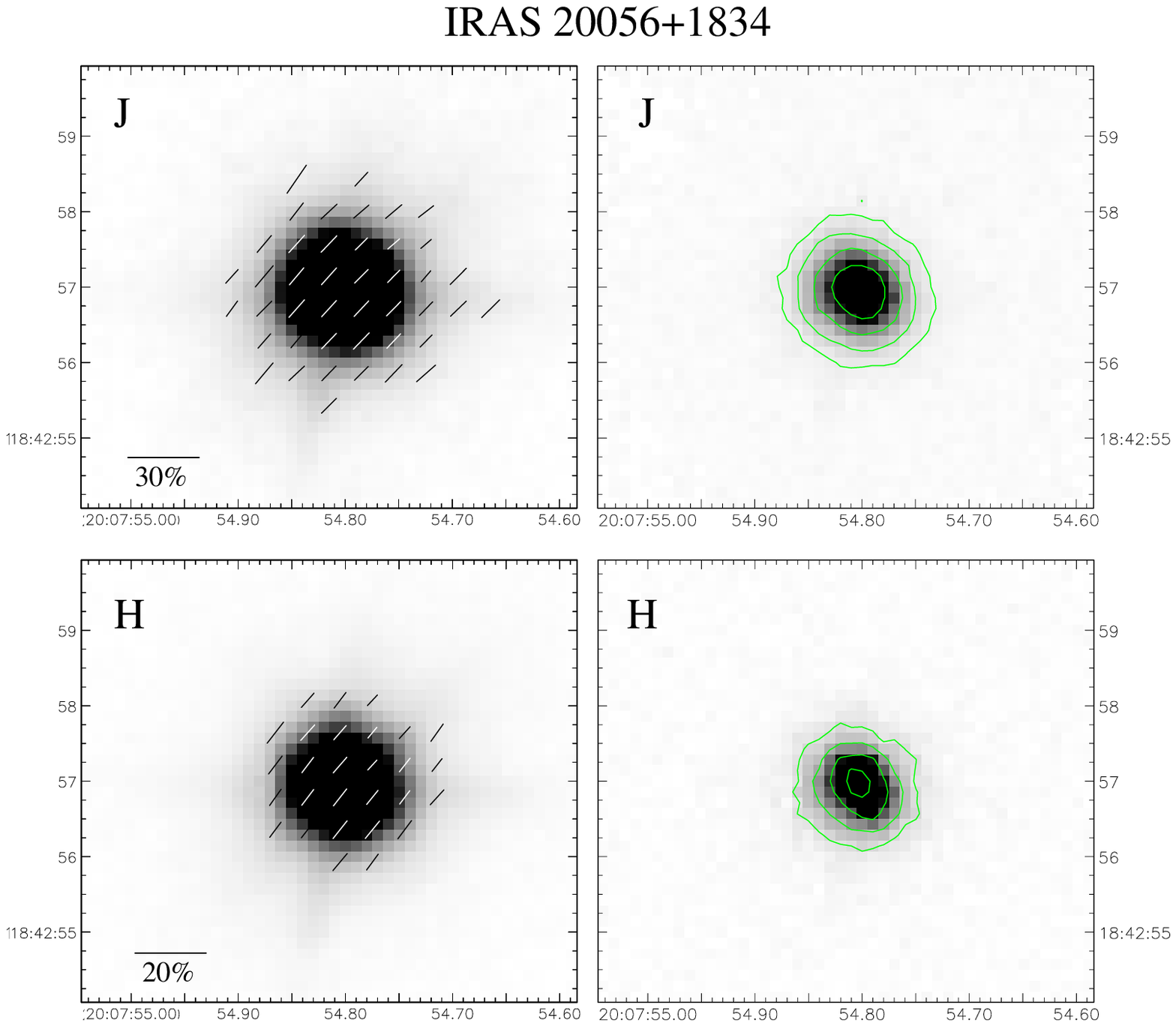}
\caption{Polarimetric observations of IRAS 20056+1834 in the J and H
bands. Details are as for Fig.~\ref{17106-fig}.}
\label{20056-fig}
\end{figure*}
The polarized flux images show none of the features typical of a
bipolar nebula. There are no well defined bipolar lobes nor dark lane
across the source. Instead, the polarized flux distribution is
centrally peaked and coincident with the peak in total flux.  However,
the polarization pattern indicates strongly anisotropic illumination
of the nebula which would be expected to result in a bipolar
appearance in polarized flux (as is seen, for example, in 17245-3951,
17106-3046, 19477+2401).  There are four other stars in the 20028+3910
field which allow us to reliably estimate the seeing at the time to be
$0.82$ arcsec (FWHM at J). We conclude then, that although the object
has a general elliptical appearance and obvious elongation, the
bipolar structure is not clearly revealed in these NIR observations.

Of the four other sources in our $\sim1\arcmin$ square field, two lie
within $6$ arcsec of 20028+3910 with the closer (bottom left in
Fig.~\ref{20028-fig}) at a separation of $3.7$ arcsec (PA
$134\degr$). It is possible, on the basis of proximity, that this star
is associated with 20028+3910, and at a level of $5\sigma_{sky}$ their
contours merge in the J band. However, there is no evidence that this
second object is polarized.

\subsubsection{IRAS 20056+1834}
\label{20056-results}

J and H band observations of 20056+1834 are shown in
Fig.~\ref{20056-fig}. In total flux there is no obvious evidence for
extension or elongation and the polarization maps show a pattern of
completely aligned vectors in both wavebands covering the region of
the source.
\begin{figure*}
\epsfxsize=18cm \epsfbox{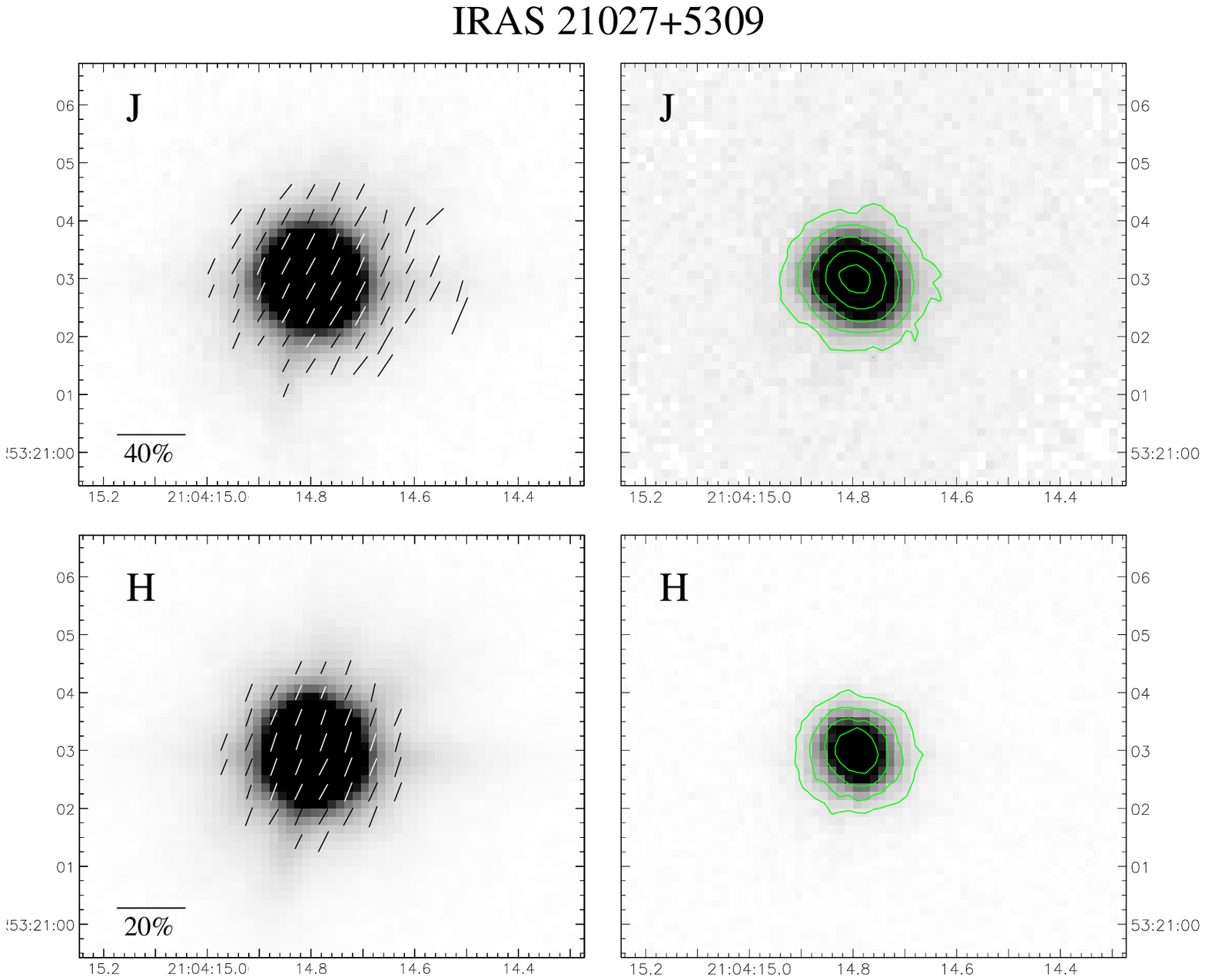}
\caption{Polarimetric observations of IRAS 21027+5309 in J and H
bands. Details are as for Fig.~\ref{17106-fig}.}
\label{21027-fig}
\end{figure*}
In polarized flux, the object appears point source-like at J and
marginally extended at H. The direction of extension at H is
perpendicular to the vector alignment angle which suggests that the
polarization may be due to scattering in an axisymmetric geometry
which is only barely resolved.

Aperture spectropolarimetry of 20056+1834 was obtained by Trammell et
al. (1994) who detected a polarization of $4.8\pm0.1$ per cent at PA
$21\pm1\degr$ (averaged between $0.5 \mu$m and $0.7 \mu$m). They find
that the PA of polarization rotates gradually with wavelength and that
the degree of polarization rises into the red. Extrapolation in
wavelength of their optical observations to the NIR leads to an
estimated degree of polarization of $\approx7$ per cent at J and
$\approx8.5$ per cent at H. This is substantially below our
measurement of $14.2$ per cent at J indicating that the polarization
increases faster than a linear extrapolation would suggest. In
addition, our observations show that the polarization decreases by a
factor of two between the J and H bands.

\subsubsection{IRAS 21027+5309}
\label{21027-results}

Observations of 21027+5309 in the J and H bands are shown in
Fig.~\ref{21027-fig}. In total flux the object appears stellar in both
wavebands and has a profile indistinguishable from that of a field
star on the same exposure. The polarization maps show a pattern of
aligned vectors across the source and appear very similar to the
results obtained for 20056+1834.  In polarized flux the object is
point source-like with no evidence for elongation or extension in our
data. The peak in polarized flux is coincident with the peak in total
flux.

Spectropolarimetry of 21027+5309 (= GL 2699) was obtained by Trammell
et al. (1994) who measured an average polarization (between $0.5 \mu$m
and $0.7 \mu$m) of $26.49\pm0.05$ per cent at PA $167\pm1\degr$.  This
is consistent with an earlier measurement of $27$ per cent and
$160\degr$ by Cohen \& Schmidt (1982). The spectropolarimetry shows a
steady decline in polarization with increasing wavelength in the
optical, which is consistent with our observed polarizations in the J
and H bands.  Polarimetrically, this object is very similar to
20056+1834 and both objects will be discussed further in
Section~\ref{core-dom-2}.

\subsubsection{IRAS 22223+4327}
\label{22223-results}
\begin{figure}
\epsfxsize=12cm \epsfbox[50 0 463 686]{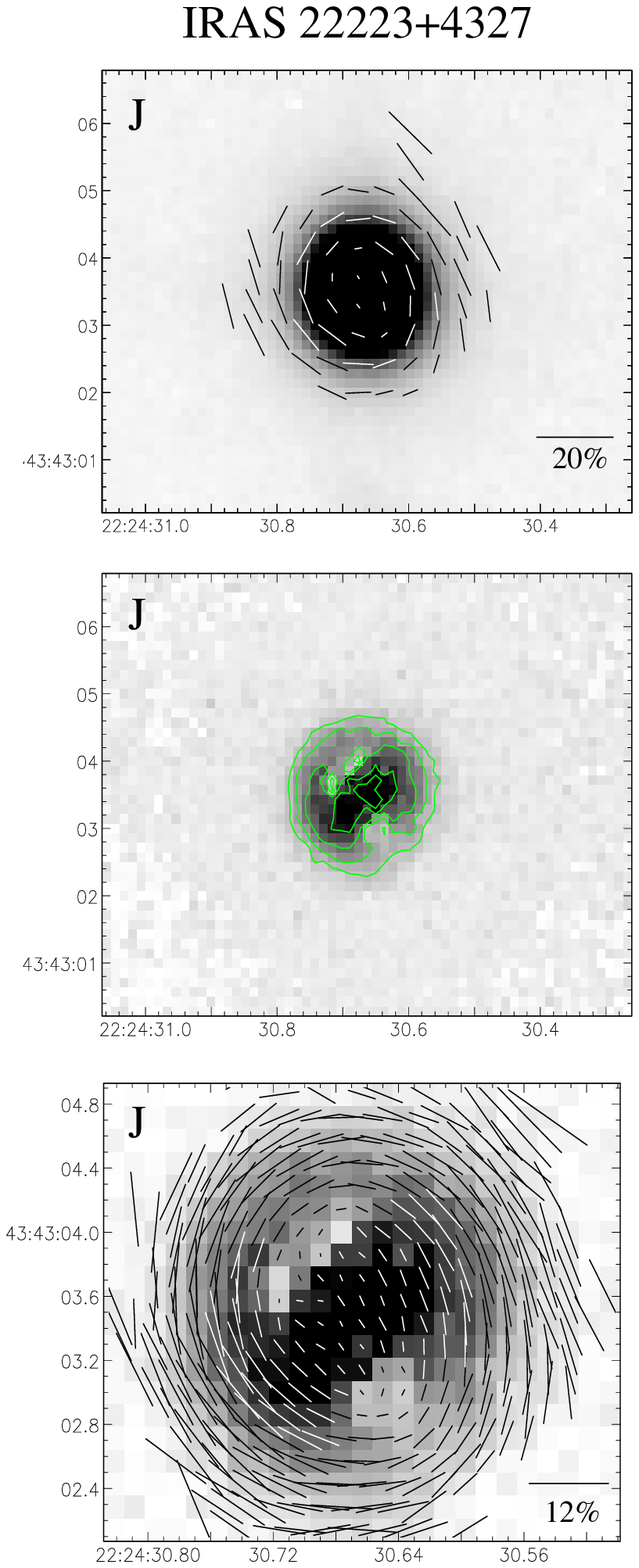}
\caption{Polarimetric observations of IRAS 22223+4327 in the J band.
The polarized flux contours are spaced logarithmically at
intervals of $0.5$ magnitudes.
Other details are as for Fig.~\ref{17245-fig} with the addition of the
lower panel showing a higher spatial resolution polarization map
($0.14\times0.14$ arcsec bins) superimposed on the polarized flux image.
 }
\label{22223-fig}
\end{figure}

Observations of 22223+4327 in the J band are shown in
Fig.~\ref{22223-fig}.  In total flux, the source appears point
source-like with no evidence for extension or elongation.  In
contrast, the polarization map clearly shows that 22223+4327 has an
extended scattering envelope that is illuminated by the central
star. The extent of the recorded polarization pattern in the J band is
$4.5$ arcsec in diameter. The vector pattern is approximately
centrosymmetric in the outer regions of the nebula but shows a slight
preference for vector alignment along a NE-SW direction. This effect
is enhanced by the lower degrees of polarization along this direction
compared with an orthogonal SE-NW axis. The polarization drops to low
levels in the core indicating that it is significantly diluted here by
the unpolarized flux from the star itself. However, significant
polarization is detected at around the $2$ per cent level throughout
the core region. A high spatial resolution polarization map (where
each vector is a single $0.143$ arcsec pixel measurement) is shown in
Figure~\ref{22223-fig} with the vector length scaled to illustrate the
polarization pattern in the core region. The vectors in the core are
aligned preferentially along a PA of $\approx40\degr$ and form a
pattern which is typical of scattering in a bipolar geometry. On
either side of the source along this axis ($40\degr$) the polarization
falls to zero marking a point where the central aligned vector pattern
changes to the centrosymmetric pattern of the outer nebula. At these
points the polarization angle rotates by $90\degr$. This effect is
often seen in studies of pre-main sequence stars (e.g. Gledhill 1991)
and can result when light is significantly polarized close to the star
(due to scattering in a bipolar geometry for example) before
scattering in a more extended envelope.

The polarized flux images of 22223+4327 are particularly interesting.
The bipolar morphology responsible for the polarization pattern in the
core region can be seen. The major axis of the bipolar nebula is at
PA $127\degr$, orthogonal to the vector pattern. Two peaks (lobes) are
discernible, offset either side of the stellar centroid. The NW peak
appears the brighter of the two suggesting a tilt out of the plane of
the sky with this lobe pointing towards us. The bipolar nebula sits within a
ring of polarized flux. The inner and outer radii of this ring are
$1.2\pm0.1$ and $2.3\pm0.1$ arcsec  respectively. 

MIR observations (Meixner et al. 1999) detected resolved emission from
22223+4327 at $12.5 \mu$m and $18.0 \mu$m ($1.9\times1.7$ arcsec at PA
$108\degr$ and $1.9\times1.8$ arcsec at PA $152\degr$ at $12.5 \mu$m
and $18.0 \mu$m respectively). These results are broadly in agreement
with ours if the MIR extension is along the axis of our
bipolar. However, it does not appear that the MIR extension
corresponds to a detection of the equatorial disc/torus responsible
for collimating the bipolar nebula, since this would require a PA in
the MIR data of $\approx40\degr$. The object is clearly axisymmetric
in our NIR data, with the star illuminating a bipolar reflection
nebula. This axisymmetry appears to coexist with an approximately
spherically symmetric shell which is also seen by reflection.

\subsubsection{IRAS 22272+5435}
\label{22272-results}
\begin{figure}
\epsfxsize=12cm \epsfbox[50 0 448 474]{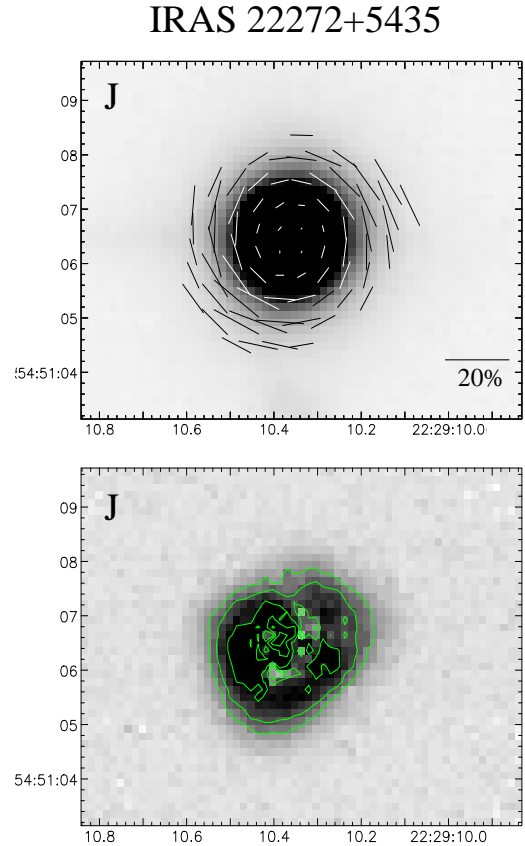}
\caption{Polarimetric observations of IRAS 22272+5435 in the J band.
The polarized flux contours are spaced logarithmically at
intervals of $0.5$ magnitudes.
Other details are as for Fig.~\ref{17245-fig}.}
\label{22272-fig}
\end{figure}

J band observations of 22272+5435 are shown in Fig.~\ref{22272-fig}.
In total flux the object shows no evidence for extension or
elongation.  The polarization map is centrosymmetric indicating that
the star illuminates an extended reflection nebula and that the
illumination is isotropic. This suggests scattering in an optically
thin environment with little evidence for axisymmetry in the dust
distribution.  The polarized flux image shows a peculiar distribution
of scattered light. A ring-like structure can be seen, with the star
located at the centre.  The ring appears brighter in the SW portion
and much fainter to the N, so that it may be incomplete. The ring is
superimposed upon a fainter halo of scattered light which has a
distinctive rectangular or box-like shape. We suggest that the
polarized flux ring corresponds to a shell of dust around the star and
that this is where most of the scattering is taking place. This shell
is part of a more diffuse envelope of dust which forms the fainter
halo. As mentioned above, on the basis of the polarimetry there is no
evidence for anisotropic illumination, indicating that dust grains in
the envelope have a direct view of the star (i.e. it is optically
thin). The polarized flux shows that the dust distribution is not
isotropic and the elongation of the faint halo along PA $127\degr$
indicates that the envelope expansion is axisymmetric. In this respect
22272+5435 is similar to 17436+5003, another object with an isotropic
illumination, a dust shell and an elongated scattering halo. This
similarity was also noted by Meixner et al. (1999) on the basis of
their MIR data.

In their optical WFPC2 imaging survey, Ueta et al. (2000) note that
22272+5435 has a bright core with four `elliptical tips'. These
elliptical tips appear to result from the superposition of two
elliptical nebulae with rotationally offset major axes. The overall
effect is to produce an outer edge to the nebula with the box-like
geometry we see in the NIR.

\subsection{Summary of Results}
\label{res-sum}
Our imaging polarimetry observations of 16 candidate PPNe show that 15
objects have extended CSEs which are seen by scattered light, in the
form of near infrared reflection nebulae. In direct (unpolarized)
light, none of the targets are obviously extended in our 4m telescope
data and many show no deviation from the PSF measured from nearby
calibration stars. It is only in scattered (polarized) light that their 
extended nature and various morphologies are revealed. We divide the
envelope morphologies into three categories, on the basis of their
polarization characteristics:
\begin{table*}
\label{res-tab}
\caption{A summary of the polarimetric results for the 16 PPNe in our
survey. We categorize the objects according to their appearance in
polarized flux as {\em shell}, {\em bipolar} or {\em core} dominated,
as described in Section~\ref{res-sum}. The important dimensions are
given along with the peak degrees of polarization (Section~\ref{P}).}
\begin{minipage}{160mm}
\begin{tabular}{p{0.7in}p{0.55in}p{0.6in}p{0.9in}p{0.7in}p{0.5in}p{1.7in}}
Target    & Category$^{a}$  & Pol. Map$^{b}$ & Polarized Flux$^{c}$ & 
     Size ($\arcsec$)$^{d}$ & PA ($\degr$)$^{e}$ & 
     Maximum Polarization (\%)$^{f}$ \\
17436+5003 & Shell   & CS   & Elongated Shell + Halo & $5.7 \times 3.6$ &
    10 & $P_{J}=20.2\pm2.2$ $P_{K}=20.0\pm4.4$ \\ 
19114+0002 & Shell   & CS   & Round Shell + Halo     & $9.0 \times 9.0$     &
    -  & $P_{J}=28.8\pm4.4$ $P_{H}=27.1\pm2.9$ $P_{K}=25.5\pm4.9$ \\
19475+3119 & Shell   & CS   & Point Symm. Shell + Halo & $4.9 \times 3.4$ &
    145 & $P_{J}=20.6\pm4.6$ $P_{K}=11.3\pm2.5$ \\
22272+5435 & Shell   & CS   & Elongated Shell + Halo & $4.0 \times 2.9$ & 
    125 & $P_{J}=18.9\pm1.4$ \\
22223+4327 & Shell/Bipolar & CS/EL & Shell + Bipolar & $2.3$ (shell) &
    127 (bip) &        \\
17106-3046 & Bipolar  & EL & Bipolar      & $3.2 \times 2.4$  & 10 &
    $P_{J}=13.0\pm4.8$ $P_{K}=6.5\pm0.8$ \\
17245-3951 & Bipolar  & AL/EL & Bipolar & $2.9 \times 1.6$ & 12 &
    $P_{J}=20.3\pm1.0$ \\
18095+2704 & Bipolar  & EL/AL & Bipolar & $2.7 \times 2.3$ & 115 &
    $P_{J}=2.4\pm0.8$ $P_{K}=2.4\pm0.3$ \\
19477+2401 & Bipolar  & EL/AL & Bipolar & $2.1 \times 0.9$ & 30 &
    $P_{J}=17.0\pm3.8$ \\
19500-1709 & Bipolar  & EL/AL & Bipolar & $3.6 \times 2.2$ & 100 &
    $P_{J}=5.7\pm1.1$ $P_{K}=6.1\pm2.1$ \\
20000+3239 & Bipolar  & EL/AL & Bipolar & $3.7 \times 3.3$ & 36 &
    $P_{J}=6.1\pm1.5$ $P_{K}=3.4\pm0.4$ \\
18184-1623 & Core     & AL    & Point Source & -           & -  &
    $P_{J}=2.4\pm0.6$ \\
20028+3910 & Core     & AL    & Elongated Halo & $4.4 \times 3.3$ & 153 &
    $P_{J}=34.0\pm2.8$ $P_{K}=41.7\pm12.5$ \\
20056+1834 & Core     & AL    & Point Source & -           & - &
    $P_{J}=14.2\pm6.1$ $P_{H}=7.4\pm1.9$ \\
21027+5309 & Core     & AL    & Point Source & -           & - &
    $P_{J}=17.6\pm3.1$ $P_{H}=6.5\pm2.8$ \\  
19454+2920 & Unpolarized & - & - & - & - & - \\
\end{tabular}
$a$ - the CSE category as defined in Section~\ref{res-sum}. \\
$b$ - the type of polarization map; CS (centrosymmetric), EL (elliptical),
AL (aligned) or a combination, as defined in Section~\ref{res-sum}. \\
$c$ - the CSE morphology in polarized flux. \\
$d$ - the extent of the CSE in our J band  polarized flux images. For 
objects in the {\em Shell} category this is the major $\times$ minor axis 
of the halo. For {\em Bipolar} objects the major $\times$ minor axes of 
the bipolar are given. For {\em Core} objects, only 20028+3910 appears 
elongated and the major $\times$ minor axes of the halo are given. \\
$e$ - The position angle (East of North) of the major axis, as defined
by the polarized flux image.\\
$f$ - The maximum polarization in each waveband after application of
various cuts (see Section~\ref{P}). Note that the errors appear large in some
cases because the largest polarizations often occur in the outer regions of 
the nebulosity where the signal to noise ratio is low. \\
\end{minipage}
\end{table*}
\begin{enumerate}
\item {\em Shells:} These objects appear shell-like in polarized flux.
Most of the scattering is taking place in a shell of dust that is
detached from the star. There are 5 objects in this category:
17436+5003, 19114+0002, 19475+3119, 22223+4327, 22272+5475. All have
centrosymmetric polarization patterns indicating that the envelope is
optically thin and illuminated isotropically by a centrally located
point source. The polarization in the core region is very low, as
expected for an optically thin shell in which the star is directly
visible. The shell is often embedded within a diffuse fainter halo of
polarized flux.

\item {\em Bipolars:} These objects are clearly bipolar in polarized
flux, with two lobes of nebulosity separated by a fainter region
across the `equator'. Objects in this category are: 17106-3046,
17245-3951, 18095+2704, 19477+2401, 19500-1709, 20000+3239,
22223+4327. The equatorial reduction in polarized flux varies from a
slight pinch in the isophotes (e.g.  17245-3952 and 18095+2704) to a
dark lane across the central region (e.g. 17106-3046,
19500-1709). These objects have non-centrosymmetric polarization
patterns. We denote the patterns as `elliptical' in that the vector
orientations define the locii of a series of ellipses, rather than a
series of circles as in the case of a centrosymmetric pattern. In the
extreme case of an `elliptical' polarization pattern (where the
`ellipticity' becomes very high) the vectors form an `aligned'
pattern. These polarization patterns are typical of scattering in
optically thick geometries where multiple scattering occurs close to
the source. They are seen frequently in observations of
pre-main sequence stars and are successfully explained using flattened
envelope and disc models (e.g. Lucas \& Roche 1998, and references
therein).

\item {\em Core Dominated:} These objects have a centrally peaked
polarized flux distribution with the peak coincident with the total
flux peak. The polarization patterns are `aligned', with the vectors
being parallel across the source. We find 4 objects in this class:
18184-1623, 20028+3910, 20056+1834, 21027+5309. In the
case of 18184-1623, 20056+1834 and 21027+5309 the polarized flux
distribution appears spherically symmetric and compact, whereas in the
case of 20028+3910 it is clearly elongated at low brightness levels
forming an elliptical halo.
\end{enumerate}

One object, 22223+4327, features in both the {\em shell} and {\em bipolar}
categories and in this respect is unique within our sample. One object,
19454+2920, appears to be unpolarized. The results are summarized in 
Table~2.

\section{Discussion}

\subsection{The Prevalence of Axisymmetry in PPNe}

In Section~\ref{res-sum} we classified our targets, on the basis of
their polarimetric properties, into three categories: {\em shell}, {\em
bipolar} and {\em core} dominated objects. We find 6 objects in the
{\em bipolar} category as well as one object (22223+4327) that
displays both {\em shell} and {\em bipolar} properties. By definition,
any object which displays a bipolar structure in polarized flux must
be axisymmetric. An indication of the degree of axisymmetry is given by 
the ratio of the extents of the major and minor axes of the bipolar and
in our data this ranges from $1.1$ for 20000+3239 to $2.5$ for 19477+2401.

There are 4 objects in the {\em shell} category (plus 22223+4327 which
is also bipolar and therefore axisymmetric) and three of these are
clearly elongated in our data. Both 17436+5003 and 22272+5435 have
elongated shells and haloes and in 19475+3119 the halo is elongated.
The axis ratios are $1.6$ for 17436+5003 and $1.4$ for 19475+3119 and
22272+5435. The fourth object in this category, 19114+0002, is less
obviously axisymmetric. Our polarized flux images
(Figure~\ref{19114-fig}) show a ring of dust surrounding the
star. Although the ring looks superficially symmetric, there are
indications of asymmetries such as inhomogeneities in the polarized
flux around the ring indicating varations in dust density. In
particular there is evidence for a thinning of the ring to the SW,
quite apparent in the K band images. Also the ring is extended
slightly to the SW (i.e. it is slightly elliptical) so that the star
is not quite at the centre.  However, there is no clear {\em axis} of
symmetry demonstrated by these features.  WFPC2 imaging observations
of 19114+0002 (Ueta et al. 2000) do show evidence for axisymmetry
though, especially the striking bipolar `protruberance' at PA
20$\degr$. As mentioned in Section~\ref{19114-results} this structure
lines up with the `gap' in our polarized flux ring. However there is
no evidence to associate the collimation of the bipolar structure with
the ring itself.  Detailed modelling of this object is currently
underway.

In the {\em core} dominated category, objects have aligned
polarization vector patterns. When accompanied by high degrees of
polarization, these indicate multiple scattering in optically thick
envelopes with an axisymmetric dust distribution. Similar patterns are
seen in nebulae illuminated by deeply embedded pre-main sequence stars
(e.g. Lucas \& Roche 1998 and references therein). In both cases (YSOs
and PPNe), the axis of symmetry in the dust envelope is perpendicular
to the vector orientation. This interpretation is supported when we
see elongation in scattered flux along this axis, such as in
20028+3910. Although there is no evidence for elongation in our images
of 20056+1834 or 21027+5309, we take their strongly aligned vector
patterns and high degrees of polarization to be evidence for
axisymmetry in these objects. In the case of 18184-1623, although an
aligned pattern is seen, the degree of polarization is comparatively
low ($P_{J}=2.4\pm0.6$ per cent). Such a low degree of polarization
could be due to line-of-sight polarization by aligned dust grains in
the ISM. However, we note that the direction of alignment is
perpendicular to the elongation of the MIR images of 18184-1623
(Robberto \& Herbst 1998) again suggesting that it is a manifestation
of axisymmetry in this source. These objects are discussed further in
the next Section.

We conclude that all of our polarized sources (15 out of 16) are
illuminating circumstellar envelopes that are to some degree
axisymmetric.  This is in accord with the surveys of Ueta et
al. (2000) and Trammell et al. (1994) who also found a preference for
axisymmetry in PPNe. It is interesting to note that the two objects
that show least evidence of axisymmetry in our survey (19114+0002 and
18184-1623) have an anomalous status. Despite sharing some of the
characteristics of PPNe, such as a double peaked SED, it is likely
that 18184-1623 is an extremely massive object in the Luminous Blue
Variable (LBV) phase, rather than a post-AGB star (Robberto \& Herbst
1998). The status of 19114+0002 is still under debate (e.g. Reddy \&
Hrivnak 1999) with the possibility that this is also a more massive
object. 

\subsection{The Nature of the Core Dominated Objects}

\subsubsection{20028+3910}
\label{core-dom-1}

Observations of 20028+3910 in the V and I bands with a spatial
resolution of $\sim0.1$ arcsec, taken with WFPC2 (Ueta et al. 2000),
show a single lobe of nebulosity in the V band, offset from the
stellar position, with a faint counterlobe emerging in the I band. The
star is not visible and appears to be totally obscured in the
optical. In addition, the relative brightness of the two lobes
suggests an inclination of the bipolar axis to the plane of the sky
such that the SE lobe is pointing towards us. Further evidence for
bipolarity is seen in the $^{12}$CO measurements of Neri et al. (1998)
which show the J=1-0 emission to be extended along the optical bipolar
axis and perpendicular to our NIR polarization vectors. In our J and K
band data, the dark lane across the star is not seen and instead of
two bipolar lobes we see a centrally peaked distribution of scattered
light, coincident with the peak in total flux. However, the aligned
polarization pattern across the source suggests scattering in an
optically thick environment and that little direct starlight is
escaping through the disc in the NIR. Detailed modelling (in progress)
is required to determine the optical depth through the disc but it is
likely that the star is only seen in scattered light in the NIR. This
is in keeping with the type II/III SED (Ueta et al. 2000; van der Veen
et al. 1989). Although no MIR photometry is available, the SED appears
to climb uniformly from the NIR to peak at around $25 \mu$m, which is
typical of other bipolar nebulae seen close to edge on, such as
17150-3224, 17441-2411 and 16342-3814. We also have NIR imaging
polarimetry data of these three bipolars (in preparation) which
(although are more obviously bipolar in appearance than 20028+3910)
show the band of aligned vectors across the source. In the case of
20028+3910, the polarization across the source, although high
($\approx23$ per cent), is lower than in the bipolar lobes
($\approx30$ per cent). This is as expected if the source region is
viewed mainly by multiply scattered light penetrating the central disc
region whereas in the lobes (which, being on the system axis have a
less obscured view of the source) single scattering is more important
resulting in higher polarization and a more centro-symmetric
polarization pattern. There is little doubt then that 20028+3910 is a
bipolar object. Its core-dominated appearance in the NIR is likely due
to an intense region of scattering close to the source.

\subsubsection{20056+1834 and 21027+4327}
\label{core-dom-2}

These objects share a number of similarities, which also indicate that
they are disparate from the rest of our sample. They were both
observed by Trammell et al. (1994) who obtained spectropolarimetry and
classified both objects as Type 1c (objects with large intrinsic
polarizations where the continua and emmission lines have different
polarizations). Our observations show aligned vector patterns across
the source and high degrees of polarization that are strongly
wavelength dependent. In both cases the degree of polarization
decreases by a factor of 2 or more between J and H
(Table~2).

What is the polarization mechanism in these objects? The lack of any
obvious scattering envelope and the uniform polarization angles would
be consistent with polarization by dichroic extinction. The
interstellar polarization of $1.23$ and $1.66$ per cent respectively
for 20056 and 21027 (Trammell et al. 1994) is much too low to account
for the source polarizations so the polarization must be intrinsic. In
a dichroic polarization interpretation, the polarization would occur
within an envelope of aligned non-spherical grains around the star.
To account for the high degrees of polarization ($P_{J}\approx17$ per
cent in the case of 21027+4327) grains would need to be coherently
aligned throughout much of the CSE. It is possible that this could be
achieved by a magnetic field permeating the envelope, but there is as
yet little evidence for grain-aligning magnetic fields around post-AGB
stars. We note, however, that the fall in polarization by a factor of
approximately 2 between the J and H bands, observed for 20056+1834, is
roughly in line with the power-law relationship derived for the dichroic
polarization of the ISM in the NIR ($P(\lambda)\propto\lambda^{-1.8}$;
e.g. Li \& Greenberg 1997). 

We favour an interpretation in which the polarization is produced by
scattering in an optically thick axisymmetric geometry, as in the case
of 20028+3910. In fact, the aligned vector patterns of 20056+1834 and
21027+5309 are just what we see in the core region of 20028+3910. We
propose that these two objects are at an earlier evolutionary phase
than 20028+3910 and are still in the process of building up a dense
equatorial dust disc. Either material has not yet been ejected along
the polar axes of these discs or polar cavities have not yet been
excavated so that we do not see a bipolar scattering nebula
perpendicular to the disc. When this does happen we would expect these
objects to develop bipolar lobes as in the case of 20028+3910.

It is of course possible that both 20056+1834 and 21027+5309 are
already bipolars with structure that has not been resolved in our
ground-based observations. However, further evidence for an earlier
evolutionary status for these objects comes from spectroscopy and
photometry which reveals that both stars are variable.  In the case of
20056+1834, the period is 50 days (Menzies \& Whitelock 1988). These
authors suggest that the star is totally obscured by an optically
thick dust cloud and is seen entirely by reflected light. On the basis
of the IR colours, which peak shortward of $10 \mu$m, they fit the
dust component of the SED with a 600 K blackbody. This is much hotter
than the typically $150-300$ K detached dust shells around PPNe and
implies that the dust is much closer to the star. This supports the
contention that this object is at an earlier evolutionary phase in
which the dust shell has not yet expanded and cooled to PPNe
temperatures. 21027+5309 (= GL2699) has been classified by Cohen \&
Hitchon (1996) as a long period Mira variable and an Extreme Carbon
Star. They derive a period of $1.34$ yr and argue that the strong
H$\alpha$ emission seen in this object results from shocking of a
stellar outflow from the inner edge of a circumstellar torus seen
almost edge-on. Trammell et al. (1994) also detect strong H$\alpha$ as
well as H$\beta$ and when Cohen and Schmidt (1982) observed this
object other lines typical of post-shock recombination (such as the
forbidden lines of OI, NII and FeII) were also present. It seems that
significant mass loss is on-going in 21027+5309 and is probably driven
by periodic shocks. This object may, therefore, still be in the
process of building up its circumstellar dust shell.

The fact that both objects show such well aligned polarization
patterns, which we attribute to scattering in an optically thick dusty
torus, provides strong evidence that axisymmetry is established before
mass loss terminates at the end of the AGB. In the case of 21027+5309,
the on-going mass loss is likely to be axisymmetric which may indicate 
that mass loss from Miras in general is axisymmetric.

\subsection{Scattering from Superwind Shells}

Molecular line studies of AGB stars, particularly CO observations,
consistently indicate that the mass loss rate on the AGB lies between
$\sim10^{-8}$ and $\sim10^{-5}$M$_{\odot}$yr$^{-1}$ (e.g. Loup et al.
1993). These observations cover hundreds of AGB stars so that the
limits seem well established and in particular imply an upper limit to
the AGB mass loss rate not exceeding a few time
$10^{-5}$M$_{\odot}$yr$^{-1}$.  This is a factor of 10 less than the
mass loss rates inferred for well-studied post-AGB objects such as
AFGL 618, AFGL 2688 and OH231.8+4.3 (see Bujarrabal 1999 and references
therein). Although there is evidence that the mass loss rate does
increase gradually throughout the AGB lifetime of a star (e.g AFGL 2688;
Sahai et al.  1998b) it is hard to escape the conclusion that the AGB
must terminate with an abrupt increase in mass loss resulting in the
ejection of a considerable bulk of material in a short time. This is
the {\em superwind} phase which, on the basis of kinematical
arguments, is expected to have a duration $t_{\small
SW}\sim10^{3}\rightarrow10^{4}$ yr (Kaufl, Renzini \& Stanghellini
1993). This dramatic increase in mass loss over a short period
(relative to the AGB lifetime) is expected to result in a {\em
superwind shell} of gas and dust, the inner edge of which marks the
cessation of mass loss and the point at which the star turns off the
AGB.

The warm dust at the inner edge of the shell, radiatively heated by
the star to $\sim200$ K, will emit at MIR wavelengths. Imaging of four
carbon-rich PPNe at $8-13 \mu$m allowed Meixner et al. (1997) to place
limits on the size of the superwind shell and, by assuming expansion
velocities from molecular line measurements, on the superwind
duration, $t_{\small SW}$. For their four PPNe, they find that the
inner radii of the shells lie in the range $0.25-1.0$ arcsec with
$t_{\small SW}$ between 800 and 3000 years. As long as the dust
optical depth is not too high then these shells should also scatter
light from the star at optical and NIR wavelengths and, with inner
radii of $\sim1.0$ arcsec, they should be resolvable in our
polarimetry data. In fact, we detect shells in five objects
(Section~\ref{res-sum}) and it is tempting to identify these
structures with superwind shells.  In the case of 17436+5003 an
elliptical shell is seen in polarized flux, superimposed on a more
extended elliptical halo. Although the centre of our image is
contaminated with polarized flux residuals resulting from the bright
star, the inner edge is well defined (especially in the J band) with a
radius of $0.6\pm0.2$ arcsec along the minor axis (PA $100\degr$).
The thickness of the shell is $1.0\pm0.2$ arcsec. The deconvolved MIR
images of 17436+5003 (Skinner et al. 1994) indicate a sharp inner
radius of the shell of $0.5$ arcsec resulting in two peaks of emission
separated by $1.0$ arcsec and orientated approximately along the minor
axis of our polarized flux image. The correspondance between the MIR
and NIR data indicate that they result from the same physical
structure and that we have detected scattering from the superwind
shell in 17436+5003. The double-peaked structure seen at $10.5 \mu$m
suggests an axisymmetric shell with an equatorial density enhancement
along PA $110\degr$ (Skinner et al. 1994). This again fits well with
our data which show an elliptical shell with major axis oriented at PA
$10\degr$ in the `polar' direction. In the J band image
(Figure~\ref{17436-fig}) the shell appears thinner in the `polar'
regions (along PA $10\degr$) and the fainter halo more extended in
this direction, suggesting that the shell is equatorially enhanced and
currently focussing the nebula expansion in the polar
direction. However, we reiterate that the polarization pattern shows
little evidence of deviation from asymmetry so that, even though the
dust density is higher at the equator then at the poles, it remains
optically thin at $1 \mu$m.

\begin{table*}
\begin{minipage}{160mm}
\label{shell-tab}
\caption{Observed properties of the shell objects. R$_{\rm{in}}$ and
R$_{\rm{sh}}$ are the sky projected inner radii and radial thicknesses
of the shells, estimated from our J band polarized flux images. Using
the object distance, D, and the expansion velocity, V$_{\rm{exp}}$
(from the literature or assumed) we calculate the duration of the superwind
phase, $t_{\small SW}$ and the time since the superwind terminated (the
dynamical time), $t_{\small DYN}$. The values of $t_{\small SW}$ and 
$t_{\small DYN}$ may be uncertain by up to 200 per cent due to uncertainties
in D and other assumptions mentioned in the text.}
\begin{tabular}{lcccccc}
Target   & R$_{in}(\arcsec)$  & R$_{\rm{sh}}(\arcsec)$ & D (kpc) &
V$_{\rm{exp}}$ (kms$^{-1}$)   & t$_{\small SW}$ (yr)  & t$_{\small DYN}$ (yr) \\
17436+5003  & $0.6\pm0.2$     & $1.0\pm0.3$            & $1.2^{a}$  &
12$^{b}$                      & 500                   & 300   \\
19114+0002  & $2.4\pm0.3$     & $1.4\pm0.3$            & $4^{c}$    &
$34^{d}$                      & 800                   & 1400  \\
19475+3119  & $0.5\pm0.2$     & $0.7\pm0.2$            & $1^{\dagger}$&
$10^{\dagger}$                & 350                   & 250   \\
22223+4327  & $0.6\pm0.2$     & $0.6\pm0.2$            & $1^{\dagger}$&
$10^{\dagger}$                & 300                   & 300   \\
22272+5435  & $0.7\pm0.1$     & $0.4\pm0.1$            & $3^{c}$    &
$9.6^{d}$                     & 650                   & 1100 \\
\end{tabular}
\\$a$: Skinner et al. (1994) \\ 
$b$: Likkel et al. (1991)  \\
$c$: Ueta  et al. (2000) \\
$d$: Zuckerman \& Dyck (1986) \\
$\dagger$: assumed value \\
\end{minipage}
\end{table*}

We have also detected shells in four other targets: 19114+0002,
19475+3119, 22223+4327 and 22272+5435. In the case of 22223+4327 the
situation is complicated since this object also has a bipolar nebula
within the shell. Also, the shell in 19475+3119 has a point symmetric
rather than axisymmetric structure (discussed further in
Section~\ref{spiral}). However, given these differences, our J band
data show that the shells are typically less than $1$ arcsec in radial
width (projected on the sky) which, assuming they result from the
superwind, can be used to constrain the duration of this
phase. Assuming an actual (rather than sky projected) angular extent
of $1$ arcsec, a distance of 1 kpc and a dust expansion velocity of
$10 $kms$^{-1}$ (a velocity typical of AGB winds, e.g. Loup et al. 1993), 
then $t_{\small SW} = 500$ yr. This value scales as:
\[
t_{\small SW} = 500 \times \frac{\rm{D}}{1 \rm{kpc}}\frac{10 \rm{kms^{-1}}}
{\rm{V_{exp}}}\frac{\rm{R}_{\rm{shell}}}{1\arcsec}~~~\rm{yr}
\]

This argument assumes that both the inner and outer edges of the shell
are detected in scattered light.  For the optically thin cases
(centrosymmetric polarization patterns) described here, detection of
the inner edge simply depends on whether it can be resolved, whereas
detection of the outer edge is a matter of sensitivity in the
observations. In a number of our objects we detect a faint outer halo
which gives us some confidence that an outer shell boundary is
seen. However, it is not certain that this represents the onset of the
superwind outflow since mass loss in the early superwind stages may
have been lower than at the end and, for an expanding shell, we expect
the surface brightness to fall of with radius even if mass loss is
constant over time. Our estimates of $t_{\small SW}$ are, therefore,
likely to be lower limits. In addition, if the equatorial dust
enhancement is sufficient for most scattering to occur in the
equatorial region (i.e. in a disc) then inclination effects will be
important in calculating the shell thicknesses from the scattered
light distribution. Having said this, the main uncertainty in
estimating shell properties in general is likely to be the distance
estimate, which may be uncertain by 200 per cent.  We list the
observed properties of our shell objects in Table~3
along with estimates of the superwind duration time, $t_{\small SW}$
and $t_{\small DYN}$, the dynamical time. The latter is determined
from the inner radius of the shell and represents the time since the
superwind terminated and during which the shell has been
expanding. For the two objects for which estimates of $t_{\small DYN}$
are available from MIR measurements (17436+5003 - $t_{\small DYN}$=240
yr [Skinner et al. 1994]; 22272+5435 - $t_{\small DYN}$=1100 yr
[Meixner et al. 1997]) the agreement with our estimates is remarkably
good, again indicating that both the MIR emission and the NIR scattered
flux originate from the same dusty shell.

\subsection{The Morphological Range of PPNe Envelopes}

We have identified a range of CSE morphologies in our observations,
which we categorize as {\em core} dominated, {\em shell} and {\em bipolar}
(Section~\ref{res-sum}) and we now discuss the possibility of
integrating these morphological types into a common model of a
post-AGB CSE. All of our objects show evidence for axisymmetry in the
envelope density and so this must be a key ingredient in any envelope
model. It is thought that mass loss for most of the lifetime of an AGB
star is spherically symmetric (e.g. Neri et al. 1998) and that
axisymmetry only emerges towards the end of the AGB. This is supported
by high spatial resolution (HST) observations of the nearby PPN AFGL
2668 (Sahai et al. 1998a,b) which show a pattern of concentric arcs on
a background of scattered light with a radial fall-off in surface
brightness, indicating that for most of the AGB lifetime the mass loss
has been spherically symmetric with occasional increases in mass loss
rate producing the arcs. This spherically symmetric shell is then
illuminated by light which appears to be collimated close to the
source in the region of the dark lane forming a highly axisymmetric or
bipolar structure.  Ueta et al. (2000) suggest that an intrinsically
axisymmetric superwind initiates the shift from a spherically
symmetric AGB envelope to an axisymmetric envelope by expelling more
material in the equatorial region. This forms a superwind shell with a
flattened density structure which is able to collimate further outflow
and expansion along the polar axes. Ueta et al. also identify two
types of object, SOLE (Star-Obvious Low-level Elongated) and DUPLEX
(DUst-Prominent Longitudinally Extended) based on the optical depth of
the envelopes. In the SOLE objects, the envelope is optically thin and
the star is visible at optical and NIR wavelengths. In the DUPLEX
objects the envelope is optically thick and the star obscured. They
argue that the two types are physically distinct, and probably result
from differing progenitor masses, rather than forming an evolutionary
sequence.

This scheme also fits our observations quite well with our {\em shell}
and {\em bipolar} categories corresponding to the SOLE and DUPLEX
categories of Ueta et al. In the case of the {\em shell} objects, the
superwind shell is optically thin in our NIR data so that we see
scattering throughout the volume of the shell. The low optical depth
means that single scattering predominates and a centrosymmetric
polarization pattern results. Increasing the optical depth will result
in the equatorial regions of the shell becoming opaque in the
optical/NIR so that the scattered light we see originates at higher
latitudes, particularly in the polar regions where the dust density is
reduced, resulting in a bipolar appearance. The increased optical
depth also results in multiple scattering leading to an elliptical or
aligned polarization vector pattern rather than a centrosymmetric
one. Therefore, it is likely that both {\em shell} and {\em bipolar}
types can be explained with the same envelope model by varying the
optical depth, and this will be investigated by detailed modelling.

The {\em core} dominated objects, in particular 21027+5309 and
20056+1834, appear to be at an earlier evolutionary phase than the
shells and bipolars, as discussed in Section~\ref{core-dom-2}. As the
CSEs of these objects expand and start to thin, we may expect them to
become more like the bipolar objects as scattered light from the polar
cavities becomes visible with decreasing optical depth and as the
angular size of the envelope increases. It is not clear, however, that
an evolutionary sequence exists between {\em core}, {\em bipolar} and
{\em shell} as the envelope evolves and expands and as the optical
depth decreases. As pointed out by Ueta et al. (2000) there is little
evidence to suggest that the PPNe with optically thin CSEs are older
than those with optically thick disc-like envelopes and bipolar
scattering lobes. In particular there is no obvious trend in the
stellar spectral types to indicate this. Therefore, we also favour the
interpretation that the optical depth of the CSE (and hence the
morphological appearance of the PPN) has more to do with the mass
outflow rate during the superwind phase than an evolutionary trend.

We do note, however, that the {\em bipolar} PPNe detected in our
survey are consistently smaller in angular extent than the {\em shell}
PPNe (Table~2). Although the distance estimates to these
objects are at best uncertain, this may indicate a younger post-AGB
status for the bipolars.
 
\subsection{Point Symmetric Structure}
\label{spiral}
\begin{figure}
\epsfxsize=12cm \epsfbox{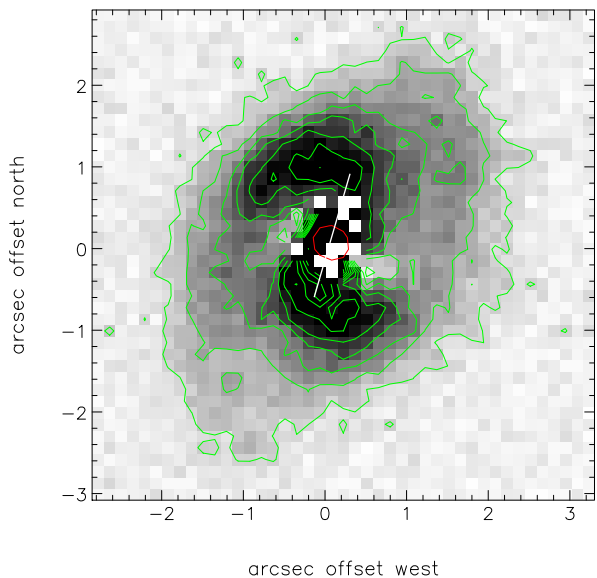}
\caption{A close-up view of the CSE of IRAS 19475+3119 seen in
scattered light (polarized flux) in the J band.  Selected contours,
spaced linearly, are superimposed on the greyscale image to highlight
the spiral arm structure. The single central contour is at 75 per cent
of the total flux peak and marks the position of the stellar
centroid. The line drawn between the two polarized flux peaks is $1.4$
arcsec in length.}
\label{spiral-fig}
\end{figure}

Point symmetric structure (PSS), that is, structure with a rotational
rather than axial symmetry, has been identified in the ionized
emission from many PNe. In a sample of 50 PNe, Gon\c{c}alves et
al. (2000) find that 27 per cent show PSS in emission lines. The
origins of PSS are currently under debate but most models involve some
form of rotating or precessing outflow, possibly involving a binary
companion (e.g. Frank 2000).  The occurence of PSS in PPNe, seen in
scattered light and molecular emission, is less well documented.

We see evidence for PSS structure in just 1 of our 15 targets with
extended CSEs. Figure~\ref{spiral-fig} shows the J band polarized flux
image of 19475+3119. The greyscale image is superimposed with selected
contours, spaced linearly in polarized flux, to highlight the spiral
structure. The stellar centroid is indicated by a single contour in
the centre at 75 per cent of the peak intensity. The two `spiral arms'
lie on either side of the star and curl around for several arcseconds
before merging with the faint elliptical halo (also see
Fig.~\ref{19475-fig}).  We have drawn a line connecting the brightness
peaks in the `arms' and this line passes through the stellar centroid
in our data, with the peaks separated by $1.4\pm0.1$ arcsec. It is
difficult to imagine how a structure with such obvious point symmetry
could be accounted for by the `standard' axisymmetric outflow and
envelope models used to explain most PPNe morphologies. Instead, an
element of rotation must be introduced. As described in
Section~\ref{19475-results}, the polarimetry indicates that the dust
responsible for scattering is optically thin in the NIR so we are
seeing the illumination of a physical structure.

One way to generate a spiral dust structure would be if an outflow
from the star is highly collimated and rotating on a timescale similar
to the outflow timescale, which in the case of the superwind would be
$\sim350$ yr (Table~3). This is a considerable departure
from the axisymmetric superwind model which posits that mass loss is
concentrated in a plane, resulting in a shell with a flattened density
distribution (e.g. Meixner et al. 1997). An alternative possibility is
that we are seeing the disruptive effects of a binary companion on the
CSE. The spiral structure bears a remarkable resemblence to the M1 and
M2 simulations of Mastrodemos \& Morris (1999) in which they treat a
binary system in which one component is an AGB star losing mass
through a spherical wind. In their model, the spiral structures are
accretion wakes resulting in patterns of density enhancement in the
orbital plane of the binary. One obvious constraint from our
observations lies in the 2-armed structure of the spiral, whereas the
simulations of Mastrodemos \& Morris show predominantly single
spirals winding around the primary. Assuming a distance of 1kpc to
19475+3119, the separation between the peaks of the spirals is
$1.4\times10^{3}$ AU, which for the R$_{p}=264$ R$_{\odot}$ primary
radius assumed in their M1 and M2 simulations, translates as
$\sim10^{3}$R$_{p}$. Hence our spirals are on a much larger scale than
the M1 and M2 structures which show spirals out to a radius of
$\sim20$R$_{p}$. Although we do not have a reliable distance estimate
for 19475+3119, our observations suggest that spiral structure can
exist on a size scale comparable with that of the CSE. Higher spatial
resolution images of this object will help to further constrain binary
models of AGB mass loss.

\section{Conclusions}


In a survey of 16 proto-planetary nebulae (PPNe), principally in the J
and K bands, we detect polarized light and image extended scattering
envelopes around 15 objects. The envelopes are typically less than $5$
arcsec in extent in our images, although one object (IRAS 19114+0002)
is larger ($9$ arcsec in diameter). In nearly all cases there is
evidence for axisymmetry in the dust density distribution, either from
the distribution of polarized flux or the pattern of the polarization
vectors (or both).  We find a range of morphological types in
polarized flux, from {\em core} dominated objects to resolved {\em
shells} and {\em bipolars}. The core dominated objects have compact
scattering envelopes and aligned polarization patterns indicative of
scattering in optically thick environments. We argue that these
objects may be at an earlier evolutionary phase so that the CSEs are
still dense, compact and close to the star. In one object there is
evidence that mass loss is still on-going. On the other hand, the
shell types are optically thin with centrosymmetric polarization
patterns with the central star visible in the NIR. We find 6 objects
with bipolar morphologies in scattered light and these generally have
`elliptical' polarization patterns indicating a higher optical depth
than in the shell-type envelopes and a degree of multiple scattering.

We find clear evidence for point symmetric structure in the CSE of
19475+3119, as seen in scattered light, in both the J and K bands. The
observed spiral structure is similar to the results of simulations 
involving a mass losing AGB star with a close binary companion 
(Mastrodemos \& Morris 1999) and so it is likely that this object
is a binary.

Apart from 19475+3119, the range of observed morphologies can be
explained with a single model of the CSE, in which a shell of dust and
gas, with an axisymmetric density distribution, is ejected late in the
AGB phase. By varying the optical depth in this `superwind shell' and
the degree to which the dust density is equatorially concentrated, a
range of scattered light morphologies will be created. In cases where
the dust shell is optically thick in the equatorial plane then
scattering in the polar cavities will be enhanced producing a bipolar
morphology. As the optical depth decreases below unity, the shell
itself will begin to scatter. 

In the 5 cases where we detect shells in
polarized flux we identify these with the `superwind shells' and use
our imaging observations to place lower limits on the duration of the
superwind phase and on the subsequent dynamical time during which the
shell has been expanding. These estimates are roughly consistent with
those from mid-infrared (MIR) observations indicating that the
scattered and thermally emitted light arises in the same physical
structure. We find that the duration of the superwind phase must have
been $\sim10^{3} yr$.


\section*{Acknowledgments}
We thank the staff of the United Kingdom Infrared Telescope which is
operated by the Joint Astronomy Centre on behalf of the U.K. Particle
Physics and Astronomy Research Council. K. Chikami is thanked for
assistance with the observations in May 1998. Data reduction was
performed at the University of Hertfordshire Starlink Node. Margaret
Meixner is thanked for comments that have improved the presentation 
of the paper.

\section*{References}
\begin{tabbing}
~~~~~\= \\
Berry D.S., Gledhill T.M., 1999, Starlink User Note 223, \\
  \> available from http://star-www.rl.ac.uk \\
Bujarrabal V., Alcolea J., Planesas P., 1992, AA, 257, 701 \\
Bujarrabal V., 1999, in Asymptotic Giant Branch Stars, \\
  \> Le Bertre T, Lebre A., Waelkens C., eds. \\
  \> IAU Symposium 191, 363 \\
Cohen M., Schmidt G.D., 1982, ApJ, 259, 693 \\
Cohen M., Hitchon K., 1996, AJ, 111 (2), 962 \\
Corradi R. L. M., Schwarz H. E., 1995, A\&A, 293, 871 \\
Frank A., 2000, in Asymmetrical Planetary Nebulae II, \\
  \> Kastner J.H., Soker N., Rappaport S., eds., \\
  \> ASP Conference Series, 199, 255 \\
Gon\c{c}alves D.R., Corradi R.M.L., Villaver E., Mampaso A., \\ 
  \> 2000, in Asymmetrical Planetary Nebulae II, \\
  \> Kastner J.H., Soker N., Rappaport S., eds., \\
  \> ASP Conference Series, 199, 217 \\
Gledhill T.M., 1991, MNRAS, 252, 138 \\
Hawkins G.W., Skinner C.J., Meixner M.M., Jernigan J.G., \\
  \> Arens J.F., Keto E., Graham J.R., 1995, ApJ, 452, 314 \\
Hrivnak B.J., Langill P.P., Su K.Y.L., Kwok S., 1999a, \\
  \> ApJ, 513, 421 \\
Hrivnak B.J., Kwok S., Su K.Y.L., 1999b, ApJ, 524, 849 \\
Jura M., Werner M.W., 1999, ApJL, 525, L113 \\ 
Kastner J.H., Weintraub D.A., 1995, ApJ, 452, 833 \\
Kaufl H.U., Renzini A., Stanghellini L., 1993, ApJ, 410, 251 \\
Kwok S., 1993, ARAA, 31, 63 \\
Kwok S., Su K.Y.L., Hrivnak B.J., 1998, ApJ, 501, L117 \\
Li A., Greenberg J.M., 1997, A\&A, 323, 566 \\
Likkel L., Forveille T., Omont A., Morris M., 1991, A\&A, 246, 153 \\
Loup C., Forveille T., Omont A., Paul J.F., 1993, A\&AS, 99, 291 \\
Lucas P.W., Roche P.F., 1998, MNRAS, 299, 699 \\ 
Mastrodemos N., Morris M., 1999, ApJ, 523, 357 \\ 
Meixner M., Skinner C.J., Graham J.R., Keto E., Jernigan J.G., \\
  \> Arens J.F., 1997, ApJ, 482, 897 \\
Meixner M., Ueta T., Dayal A., Hora J.L., Fazio G., \\
  \> Hrivnak B.J., Skinner, C.J., Hoffmann, W.F., \\
  \> Deutsch L.K., 1999, ApJS, 122, 221 \\
Menzies J.W., Whitelock P.A., 1988, MNRAS, 233, 697 \\
Neri R., Kahane C., Lucas R., Bujarrabal V., Loup C., \\
  \> 1998, A\&AS, 130, 1 \\ 
Reddy B.E., Hrivnak B.J., 1999, AJ, 117, 1834 \\
Robberto M. \& Herbst T.M., 1998, ApJ, 498, 400 \\
Sahai R., Hines D.C., Kastner J.H., Weintraub D.A., \\
  \> Trauger J.T., Rieke M.J., Thompson R.I., \\
  \> Schneider G., 1998a, ApJ, 492, L163 \\
Sahai R., Trauger J.T., Watson A.M., Stapelfeld K.R., \\
  \> Hester J.J., Burrows C.J., Ballister G.E., Clarke J.T., \\
  \> Crisp D., Evans R.W., et al., 1998b, ApJ, 493, 301 \\
Skinner C.J., Meixner M., Hawkins G.W., Keto E., \\
  \> Jernigan J.G., Arens J.F., 1994, ApJ, 423, L135 \\
Su K.Y.L., Volk K., Kwok S., Hrivnak B.J., 1998, ApJ, 508, 744 \\
Trammell S.R., Dinerstein H.L., Goodrich R.W., \\
  \> 1994, AJ, 108 (3), 984 \\
Ueta T., Meixner M., Bobrowsky M., 2000, ApJ, 528, 861 \\
van der Veen W.E.C.J., Habing H.J., Geballe T.R., 1989, \\
  \> A\&A, 226, 108 \\
Volk K., Kwok S., 1989, ApJ, 342, 345 \\
Zuckerman B., Dyck H.M., Claussen M.J., 1986, ApJ, 304, 345 \\
\end{tabbing}

\end{document}